\documentclass[twocolumn]{aastex6}

 \usepackage{appendix}
\usepackage{graphicx}
\usepackage[percent]{overpic}
\usepackage{txfonts}
\usepackage{xcolor}
\usepackage{pict2e}
\usepackage{pbox}
\usepackage{booktabs}
\usepackage{multirow}
\usepackage{color, colortbl}
\definecolor{Gray}{gray}{0.9}
\usepackage{placeins}
\begin{document}

\title{Solar Flare Prediction Using Magnetic Field Diagnostics Above the Photosphere}

\author{M. B. Kors\'os\altaffilmark{1,2}, M. K. Georgoulis\altaffilmark{3,4}, N. Gyenge\altaffilmark{5}, S. K. Bisoi\altaffilmark{6},  S. Yu\altaffilmark{7}, S. Poedts\altaffilmark{8,9}, C. J. Nelson\altaffilmark{10}, J. Liu\altaffilmark{11}, Y. Yan\altaffilmark{6} and R. Erd\'elyi\altaffilmark{11,2} }

\altaffiltext{1}{Department of Physics, Aberystwyth University, Ceredigion, Cymru, SY23 3BZ, UK}
\altaffiltext{2}{Department of Astronomy, E\"otv\"os Lor\'and University, P\'azm\'any P\'eter s\'et\'any 1/A, H-1112 Budapest, Hungary}
\altaffiltext{3}{Department of Physics \& Astronomy, Georgia State University, Atlanta, GA 30303, USA}
\altaffiltext{4}{RCAAM of the Academy of Athens, 11527 Athens, Greece}
\altaffiltext{5}{Research Computing Support, IT services, The University of Sheffield, 10-12 Brunswick Street, Sheffield, S10 2FN, UK}
\altaffiltext{6}{Key Laboratory of Solar Activity, National Astronomical Observatories, Chinese Academy of Sciences, Beijing 100100, China}
\altaffiltext{7}{New Jersey Institute of Technology, Center for Solar-Terrestrial Research, 323 Martin Luther King Blvd Newark, NJ 07102, United States}
\altaffiltext{8}{Dept. Mathematics/Centre for mathematical Plasma Astrophysics, KU Leuven, Celestijnenlaan 200B, 3001 Leuven, Belgium }
\altaffiltext{9}{Institute of Physics, University of Maria Curie-Sk{\l}odowska, Lublin, Poland}
\altaffiltext{10}{Astrophysics Research Centre (ARC), School of Mathematics and Physics, Queen’s University, Belfast, BT7 1NN, NI, UK}
\altaffiltext{11}{Solar Physics \& Space Plasma Research Center (SP2RC), School of Mathematics and Statistics, University of Sheffield, Hounsfield Road, S3 7RH, UK}

\email{robertus@sheffield.ac.uk}

\begin{abstract}
{In this article, we present the application of the weighted horizontal gradient of magnetic field ($WG_{M}$) flare prediction method to 3-dimensional (3D) extrapolated magnetic configurations of 13 flaring solar active regions (ARs). 
The main aim is to identify {\it an optimal height range}, if any, in the interface region between the photosphere and lower corona, where the flare onset time prediction capability of $WG_{M}$ is best exploited. The {\it optimal height} is where flare prediction, by means of the $WG_{M}$ method, is achieved earlier than at the photospheric level.
3D magnetic structures, based on potential and non-linear force-free field extrapolations, are constructed to study a vertical range from the photosphere up to the low corona with a 45 km step size. The $WG_{M}$ method is applied as a function of height to all 13 flaring AR cases that are subject to certain selection criteria. We found that applying the $WG_{M}$ method between 1000 and 1800 km above the solar surface would improve the prediction of the flare onset time by around 2-8 hours.Certain caveats and an outlook for future work along these lines are also discussed.}

\end{abstract}


\section{Introduction}

The short-term (i.e., hours to days) interaction of solar activity manifestations with geospace occurs through a complex series of events, commonly referred to as Space Weather (SW). Solar activity contributing to SW generally falls under one of four major components: solar flares, coronal mass ejections (CMEs), high-speed solar wind and solar energetic particles. From these occurrences, two most prominent ones are, arguably, solar flares and CME eruptions \citep{Schwenn2006}.
Earth is always impacted by Earth-facing solar flares, with impacts increasing with flare size. Major flares can generate long-lasting radiation storms in the Earth's upper atmosphere causing serious radio or data communication blackouts, amongst other damaging effects. Flares of larger GOES classes are more frequently associated with CMEs \citep[see, for example,][]{Yashiro2005}. 

CMEs, however, can be even more hazardous than flares. They are large clouds of magnetised plasma that may plow right through the Sun-Earth interplanetary space at high speeds. The impact of CMEs on Earth's magnetosphere can influence or even damage a number of socio-economically vital ground-based (e.g.\ long-distance oil or gas pipelines, electric power networks) and space-borne (satellites for communication, navigation (GPS, ISS,...) infrastructures \citep{Eastwood2017}. 
Many of these societal assets and services are key to the global economy, security and wellbeing. Considerable infrastructure failures by CMEs have indeed happened in the past (e.g. the March 1989 electrical power blackout in Quebec, Canada). The largest known and potentially most dangerous solar eruption in recent history avoided Earth by only $\sim$30 degrees in 2012 \citep{Temmer2015}. 

The frequency of occurrence of these most energetic eruptions in the entire Solar System follows the 11-year solar cycle. At the peak of the cycle, intense flares and powerful CMEs occur frequently (i.e. around 2-3 daily). It is widely accepted that major solar eruptions (i.e., flares and CMEs, or eruptive flares) originate mostly from magnetically complex, highly twisted and sheared elements of an active region (AR), typically around sunspot groups with mixed magnetic polarities (called $\delta$-sunspots) \citep[e.g.][]{Toriumi2019, Georgoulis2019}.
A key direction of research in solar eruptive activity aims to understand the dynamics of $\delta$-sunspots preceding flare and CME eruptions in order to predict these eruptions within practical timescales, enabling protection of our high-tech facilities and, of course, ourselves. Predicting reliably and accurately these solar eruptions is a major scientific endeavour on its own. The question is not ``whether" but ``when" a potentially devastating flare (or CME) may happen, with adverse effects on our technosphere. 

There are a number of methods available in the literature that rely on a range of predictive parameters of solar eruptions \citep[see e.g.][and references therein]{Barnes2016,Leka2019}. Most flare and CME forecast methods apply photospheric magnetic and Doppler data of ARs for forecasting. Some recent, pioneering approaches with various degrees of success attempt to incorporate solar atmospheric extreme ultraviolet (EUV) data and /or use Machine Learning in order to improve forecasting accuracy \citep[see e.g.][]{Qahwaji2007,Bobra2015,Florios2018,Kim2019, Yimin2019,Campi2019}. 
Detailed information on measuring, and the consequent modelling, of the 3D magnetic field structure of an AR would be important to obtain more accurate insight into the pre-flare evolution locally in the solar atmosphere. However, direct routine observations of the 3D magnetic field in the lower solar atmosphere, above the photosphere up into the top of the chromosphere, are currently not available, with an overwhelming majority of observations referring now to either the line-of-sight (LOS) component or the full magnetic field vector in the photosphere.
Nowadays, approximate methods for modelling the local magnetic field vector in the solar atmosphere include its construction using current free (potential, PF) or nonlinear force-free field (NLFFF) extrapolation techniques. In practice, however, to construct an accurate and reliable 3D magnetic field structure of an AR from photospheric measurements is still a challenging task with a number of caveats, see e.g. \cite{Wiegelmann12}.

Another potentially insightful approach may be the numerical simulation of AR from the sub-photosphere to their emergence and evolution in the lower solar atmosphere. With the aims of testing flare prediction with simulated data, a flaring AR with $\delta$-sunspots was modelled by \cite{Korsos2018a}. They introduced and applied two flare precursors, part of the $WG_M$ method \citep{Korsos2019}: one is related to the ``inverted V-shape'' feature of the $WG_M$ proxy and the other is obtained from the ``U-shape'' of the so-called distance parameter prior to each investigated flare at a certain height range in the solar atmosphere. \cite{Korsos2018a} further conjectured the existence of the so-called {\it optimal height}, where the ``U-shape'' manifests itself earlier and reaches its minimum value earlier than in the photosphere. 
In their modelling study it was also shown that these optimal heights agreed reasonably well with the heights of flare occurrence identified by an analysis of thermal and Ohmic heating signatures enabled by the magnetohydrodynamic (MHD) simulations of \cite{Korsos2018a}. 
Next, for NOAA AR 11429, \cite{Korsos2018b} used PF extrapolation to construct the 3D magnetic field above the photosphere and studied the pre-flare evolution of this AR prior to two M-class flares. There, it was found again, that the earliest onset time estimation was enabled at a distinct and specific height range, i.e.\ at an {\it optimal height}, when compared to patterns derived from data in the photosphere or other atmospheric heights.

The two above-mentioned studies prompt us to attempt to further establish the details of the conjectured wide applicability and benefits of 3D pre-flare analyses using a larger sample of flaring ARs and seeking the relevant {\it optimal height(s)} with better statistical significance. 
This work is organised as follows: Section~\ref{selection} describes the adopted tools for the pre-flare analysis of a given 3D solar magnetic structure. Section~\ref{casestudy} introduces and describes the application of the $WG_{M}$ to the lower solar atmosphere before the flare occurrences. Section~\ref{statistic} discusses our results in detail.  We summarise our key findings and draw our conclusions in Section~\ref{conclusion}. The Appendix contains more cases that support the analysis presented in the main body of the work.

 \section{Methodology} \label{selection}

\begin{figure}[h!]
\centering
\includegraphics[width=0.49\textwidth]{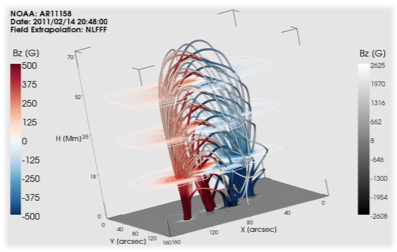}
\caption{\label{3D11158} Three-dimensional NLFFF extrapolation of  AR 11158 at 20:48:00 on 14/02/2011. The red-blue colour bar refers to the positive and negative polarity magnetic field values at different heights in the solar atmosphere. The grey colour bar represents the photospheric vertical magnetic field component, B$_{z}$. The three horizontal slices represent the identified sunspot at various heights in the lower solar atmosphere.}
\end{figure}

Motivated by the case studies of \cite{Korsos2018a,Korsos2018b}, we now extend the application of the $WG_{M}$ method to more ARs by constructing a data catalogue of sunspots using 3D PF and NLFFF extrapolations. 
An extrapolation example from the collected data catalogue is shown in Figure~\ref{3D11158}. 

With extrapolations accomplished, the $WG_{M}$ method is applied to the 3D PF and 
NLFFF extrapolated data. The $WG_{M}$ method is then applied to both data sets and the results are compared. 

It is clear that each of the two extrapolation types, PF and NLFFF, has its strengths and weaknesses: the NLFFF is most likely a much better reconstruction method of the magnetic field in the lower solar atmosphere of an AR than the PF, it is considerably more expensive computationally. Hence, if the advantage in using NLFFF against PF toward improving the lead time for flare onset prediction is trivial, one might opt to save computing time by using the PF approximation instead of the NLFFF one.

\FloatBarrier
\subsection{Selection of studied ARs} \label{3D}

Before we initiate the 3D analysis of the pre-flare dynamics of ARs with the $WG_{M}$ method, a consistent dataset of ARs is required. The data processing and the PF/NLFFF magnetic field extrapolations of ARs were carried out with the extensive use of SolarSoft\footnote{http://www.mssl.ucl.ac.uk/surf/sswdoc/solarsoft/ssw\_install\_howto.html}, with purpose- and instrument-specific routines. 
For an AR to be included in the analysis, the following four selection criteria are set to be satisfied: 
\begin{enumerate}
\item The studied AR is located between $-60^{\circ}$ and $+60^{\circ}$ in central meridian distance during the examined period of time. 
\item The AR hosted a GOES X-class flare. This is dictated by practical, computational reasons and can be revisited when sufficient resources are available.
\item The easternmost central meridian distance of X-flaring locations is not more than $\sim -40^{\circ}$.
\item The AR had at least one $\delta$-spot(s). 
\end{enumerate}

Over Solar Cycle 24, 13 ARs were found to satisfy the above four selection criteria.

\subsection{3D lower atmospheric magnetic field of ARs}

Both the PF and the NLFFF extrapolations require photospheric boundary conditions. We employed the Solar Dynamics Observatory Helioseismic and Magnetic Imager \citep[HMI,][]{Scherrer2012} LOS magnetograms as boundary condition for the PF extrapolation. For the NLFFF, the vector magnetic field measurements of HMI Active Region Patches (HARP) \citep{Bobra2014} are used as a boundary condition. In this work, the magnetogram data were studied every hour and were resized by a factor of 8 thus giving rise to photospheric magnetograms with a pixel size of 4 arcsec.

\subsubsection{PF extrapolation}
To determine the magnetic field above the photosphere with the PF extrapolation method, we employed the linear force free field (LFFF) IDL extrapolation code (see \url{www.heliodocs.com}), that is based on \cite{Gary1989}, where we set the force-free parameter ($\alpha$) equal to zero.

In brief, the PF is the simplest possible assumption for the solar atmospheric magnetic field. The LOS magnetograph is used as boundary condition to solve Laplace's equation,

\begin{equation}	
       \nabla^{2}\phi=0,
        \label{Helmholtz}
\end{equation}	

where $\phi$ is the associated scalar potential for the PF.

\subsubsection{NLFFF extrapolation}

We apply the direct boundary integral formulation for nonlinear force-free magnetic field extrapolation as outlined by \cite{Yan2006}. 
The predecessor of the NLFFF extrapolation used here is the so-called boundary integral method, first developed by \cite{Yan2000}, and recently implemented with GPU-acceleration by \cite{Wang2013}. 

The method uses the Green's function to reformulate the NLFFF problem. The obtained nonlinear integral equations allow the independent calculation of the vector magnetic field at any location of the extrapolation volume. The method considers the half-space above the lower boundary with vanishing magnetic field at infinity.  
 The solution at a given point $i$ inside the volume $V$ for the boundary magnetic field values ({\bf B}$_{0}$) on $\Gamma= \partial$V, is given by:

 \begin{equation}	
  c_{i}{\bf B}_{i} =   \oint_{\Gamma} { \Bigg( {\bf Y} \frac {\partial {\bf B}}  {\partial n}  -  \frac {\partial {\bf Y}}  {\partial n}  {\bf B}_{0}  \Bigg) } d \Gamma,
    \label{boundary}
\end{equation}	
with $c_{i}$=1 for points in the volume and  $c_{i}$=1/2 for boundary points. {\bf Y}, in Eq.~(\ref{boundary}), is a kernel function which depends on {\bf B} \citep[see for more details the Eq. (19) in][]{Yan2000}.

 \subsection{Catalogue}

We tracked ``sunspots'' above their photospheric altitudes using the Yet Another Feature Tracking Algorithm \cite[YAFTA;][]{Welsch2003, DeForest2007}. YAFTA is accessible from the Solarsoft IDL library. The detection algorithm is based on the so-called clumping method, which enrols together all contiguous-like-polarity pixels with absolute flux densities above a specified threshold, and marks them as unique element. In our study, YAFTA is grouping pixels into an element (such as a sunspot) when the given criteria are satisfied: i) minimum number of pixels is 30 and ii) local vertical magnetic fields exceed a $\left|150\right|$ G threshold.

The magnetic field strength, area, and cross-sectional diameter of all identified sunspots of ARs are then recorded for each relevant frame and saved in the sunspot data catalogue. 
The generated 3D catalogue includes the area, mean magnetic field and location (Carrington coordinates, {\it L} and {\it B}) of identified sunspots using a 45 km step above the photosphere toward the lower corona. The 45 km step size was chosen as the highest vertical resolution, i.e. smallest grid size, implemented in the NLFFF extrapolations.

\subsection{Identification of $\delta$-spot of an AR} \label{PILaut}

\begin{figure}[h!]
\centering
\includegraphics[width=0.49\textwidth]{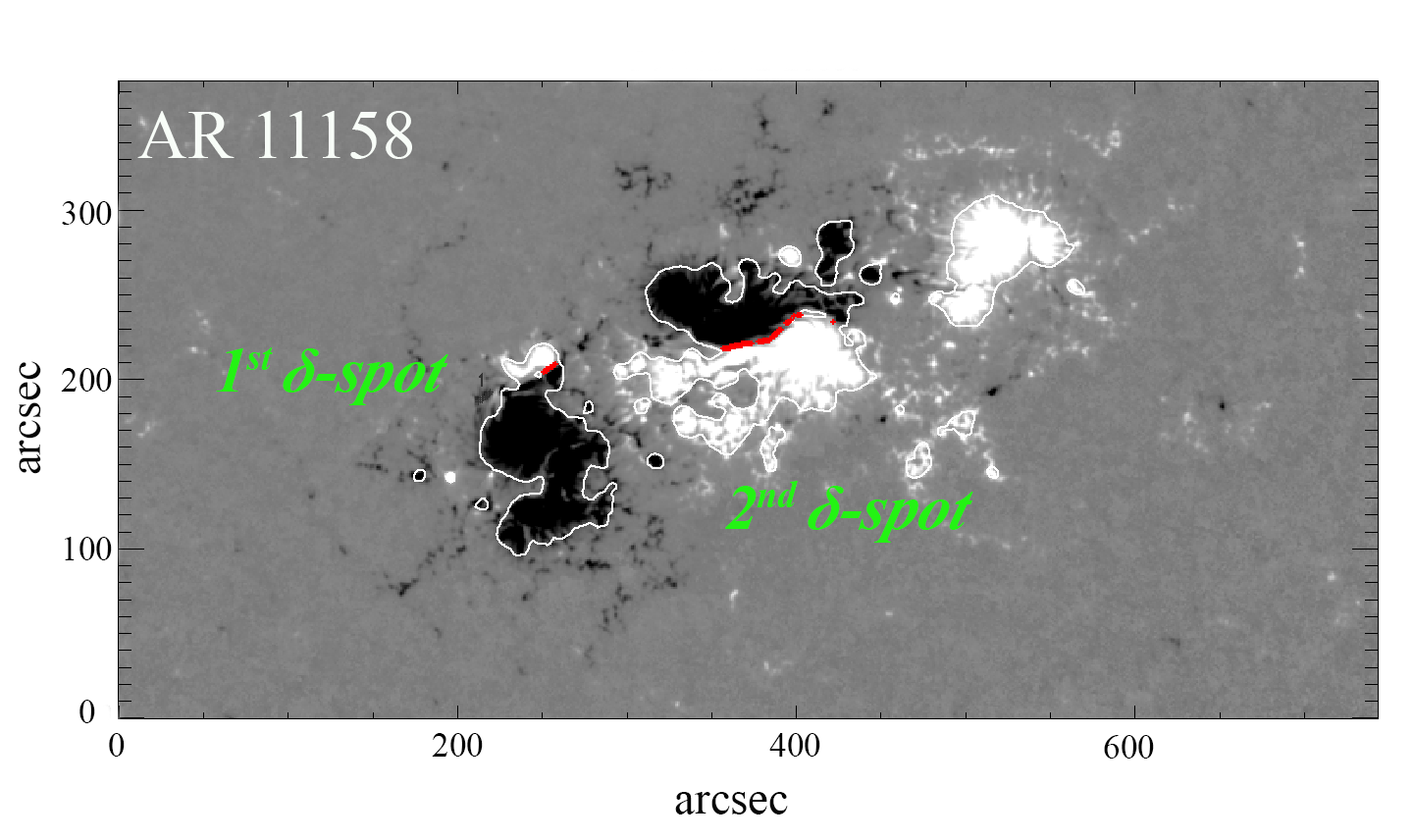}
\caption{\label{PIL} Magnetogram snapshot showing two $\delta$-spots in NOAA AR11158 on 14/02/2011 at 20:48:00. The red dotted lines are the automatically identified PILs of the AR. White contours define areas that enclose the identified strong flux elements.}
\end{figure}

Before we begin to apply the $WG_{M}$ method, we need to identify the $\delta$-spot(s) of the selected ARs. Here, we adopt and employ the automatic PIL recognition algorithm developed by \cite{Cui2006}. The program first computes the horizontal component of the PF. Next, the pixels are selected, based on whether the strength of the deduced transverse component of the magnetic field is higher than $\left|150\right|$ G. Also, the pixels are identified where the horizontal gradient of the longitudinal component of the magnetic field is larger than $\left|50\right|$ G/Mm. 
In the example given in Fig.~\ref{PIL}, the contoured area with PIL(s) corresponds to the  $\delta$-spots in the region, where we apply the $WG_M$ method.

\section{Analysis of pre-flare behaviour based on extrapolation data} \label{casestudy}

Through the case of AR 11158 with an X2.2 flare at 01:56 on 15/02/2011, let us now demonstrate the application of the $WG_{M}$ method as a function of height, applying it to PF and NLFFF extrapolations. AR 11158 has two $\delta$-spots, labelled as the 1$^{st}$ and 2$^{nd}$ $\delta$-spots, see in Fig.~\ref{PIL}.

For both of the PF and NLFFF extrapolation results, let us now track the evolution of (i) the $WG_{M}$ proxy, (ii) the $D_{pn}$ distance between the area-weighted barycenters of opposite polarities and (iii) the  unsigned magnetic flux $\Phi$ in the 1$^{st}$ and 2$^{nd}$ $\delta$-spots at consecutive 45 km steps in height ($z$). To identify the ``inverted V-'' and ``U-shape" pre-flare features, we use the maximum and minimum values of the best {\it n}th degree polynomial fit to the $WG_{M}$ and $D_{pn}$ data, respectively.

We, hereafter, focus on those atmospheric heights, where the ``inverted V-'' and ``U-shape" are identifiable prior to the X2.2 flare, for both $\delta$-spot cases. For NOAA AR 11158 we have found the following:
\begin{itemize}

\item In the case of the 1$^{st}$ $\delta$-spot, the ``inverted V-shape'' of the $WG_{M}$ and the ``U-shape" of the $D_{pn}$ are not discernible in the two extrapolation cases (see e.g. in App. Fig.~\ref{fig11158f}a-b).

\item In the case of the 2$^{nd}$ $\delta$-spot, the ``inverted V-'' and ``U-shape" each are observed prior to the X2.2 flare in both extrapolation cases (e.g. see Figs.~\ref{fig11158s}-\ref{fig11158t} a-b). These two pre-flare behaviours concurrently manifest, from the photosphere up to 3000 km in the low corona. 
Also, this $\delta$-spot was the actual major source region of the X2.2 flare \citep[e.g.][]{Wang2012}.
\end{itemize}

After, the relevant ``inverted V-'' and ``U-shape" are successfully identified at a given height, we carry on investigating their evolution as function of height. In particular, based on \cite{Korsos2018a,Korsos2018b}, we concentrate on the starting (first orange dots in Figures~\ref{fig11158f}--~\ref{fig11158t}) and finishing (first blue dots in Figs.~\ref{fig11158f}-~\ref{fig11158t}) moments of the converging phase of the  ``U-shape" illustrated at various heights/cases, because, the elapsed time between these two moments provides information about the expected flare onset time \citep[see for more details Fig. 5 of][]{Korsos2019}.

\begin{figure}[]
\centering
\includegraphics[width=0.49\textwidth]{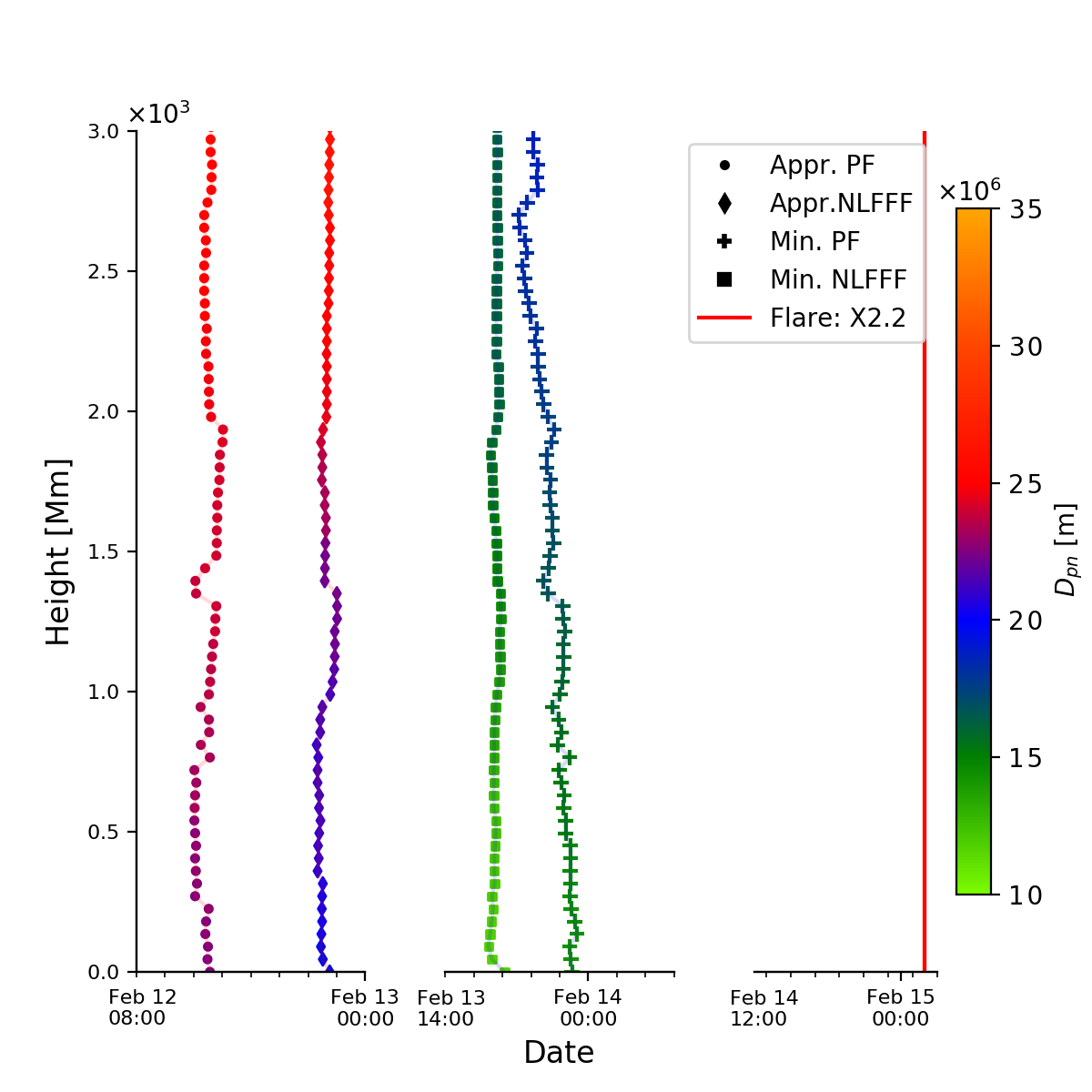}
\put(-150,225){\scriptsize 2$^{nd}$ $\delta$-spot}
\put(-201,225){\scriptsize 2$^{nd}$ $\delta$-spot}

\caption{\label{11158height} {\it AR 11158}: Times associated with the start (point/diamond symbols) and closest (plus/square symbols) convergence of opposite-polarity area-weighted barycenters as a function of height. Results obtained using the PF and the NLFFF extrapolations are shown by point/plus and diamond/square symbols, respectively. The moments of starting and closest convergence times are deduced by the maximum and minimum values, respectively, of the best {\it n}th degree polynomial fit to the $D_{pn}$ data. The colour bar gives information about the actual value of $D_{pn}$. Results from 2$^{nd}$ $\delta$-spot in the AR are shown.The red vertical line marks the X2.2 flare occurrence time that occurred from the 2$^{nd}$ $\delta$-spot. }
\end{figure}

Figure~\ref{11158height} shows the starting time ($D_{pn}^{Max}$, point and diamond symbols) and finishing time ($D_{pn}^{Min}$, plus and square symbols) of the converging phase at each 45 km step in the two $\delta$-spot cases.
The point/plus symbols represent the data derived from the PF extrapolation and the diamond/square symbols indicate the results of the NLFFF extrapolation for the constructed 3D lower atmospheric magnetic fields of AR 11158. The colour code corresponds to the actual value of the $D_{pn}$.  Also, the red line marks the onset time of the X2.2 flare in AR 11158. We notice that {\it the converging phase begins earlier and reaches its minimum distance also earlier at a certain height} (referred to as the {\it optimal height}) than it does at the photosphere, as also found in \cite{Korsos2018a,Korsos2018b}. 

Identifying the corresponding optimal heights, we estimate the expected largest flare intensity class ($S_{flare}$) and onset time ($T_{est}$), as in \cite{Korsos2019}, to investigate the applicability of $WG_{M}$ for 2$^{nd}$ $\delta$-spot:
\begin{itemize}
\item In the PF case, the optimal height is 1395 km because the converging phase (the max point of the fitted {\it n}th order polynomial) started 1.2 hrs before and finished 2 hrs earlier than in the photosphere, enabling the maximum lead time prior to flare onset as function of solar atmospheric height.
\item In the NLFFF case, the converging phase began 0.9 hrs before and finished 0.7 hrs earlier at best, at 810 km. 
\end{itemize}
Therefore, with the PF approximation one can estimate the expected flare onset time 1.3 hr earlier than in the case of NLFFF. It also worth noting that $T_{est}$ is very close to the actual values of $T_{D+F}$ in the case of NLFFF.
At these two optimal heights, the $S_{flare}$ of the investigated flare are found to be fairly well estimated, i.e. the expected flare intensity is determined as an X-class flare (see Table~\ref{1table}).

The above obtained two estimates ($S_{flare}$ and $T_{est}$) are summarised in Table~\ref{1table}. Furthermore, Table~\ref{1table} includes information about the time prior to the flare at the start ($T^{C}_{lmp}$) and closest ($T^{M}_{lmp}$) convergence at the optimal height, for both extrapolation approaches. $T^{M}_{lmp}$ can also be understood as the {\it lead-time} at the corresponding optimal height, as we estimate the flare onset time from $D_{pn}^{Min}$-$D_{pn}^{Max}$. Actually, the values of $T_{Imp}^{M}$ indicate how much time one could gain in the flare onset time estimation if one applies the $WG_{M}$ method at an identified optimal height.

\subsection{Additional results of PF vs. NLFFF}\label{result}

\begin{figure}[]
\centering
\includegraphics[width=0.49\textwidth]{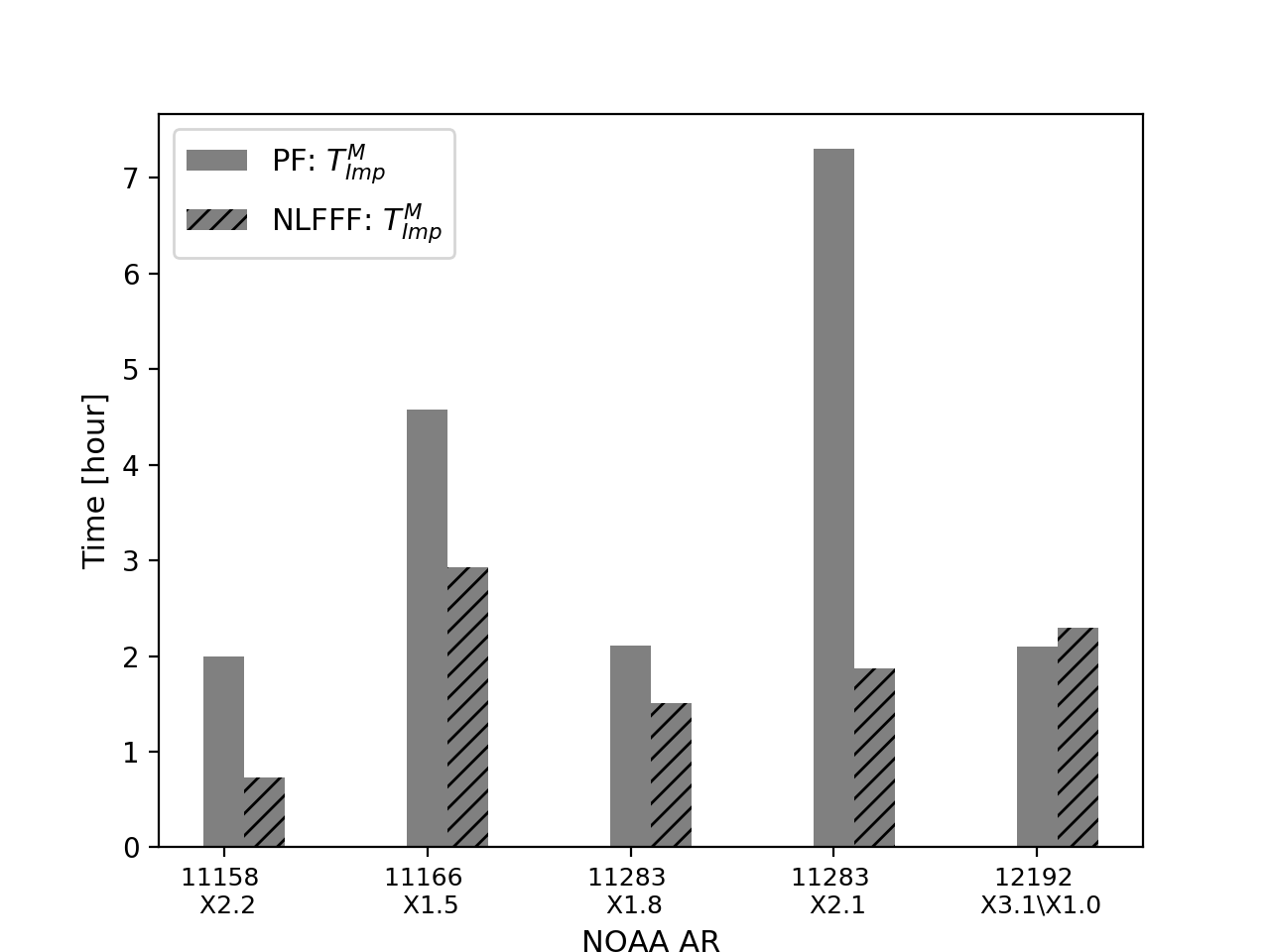}
\put(-240,170){(a)}
\\
\includegraphics[width=0.49\textwidth]{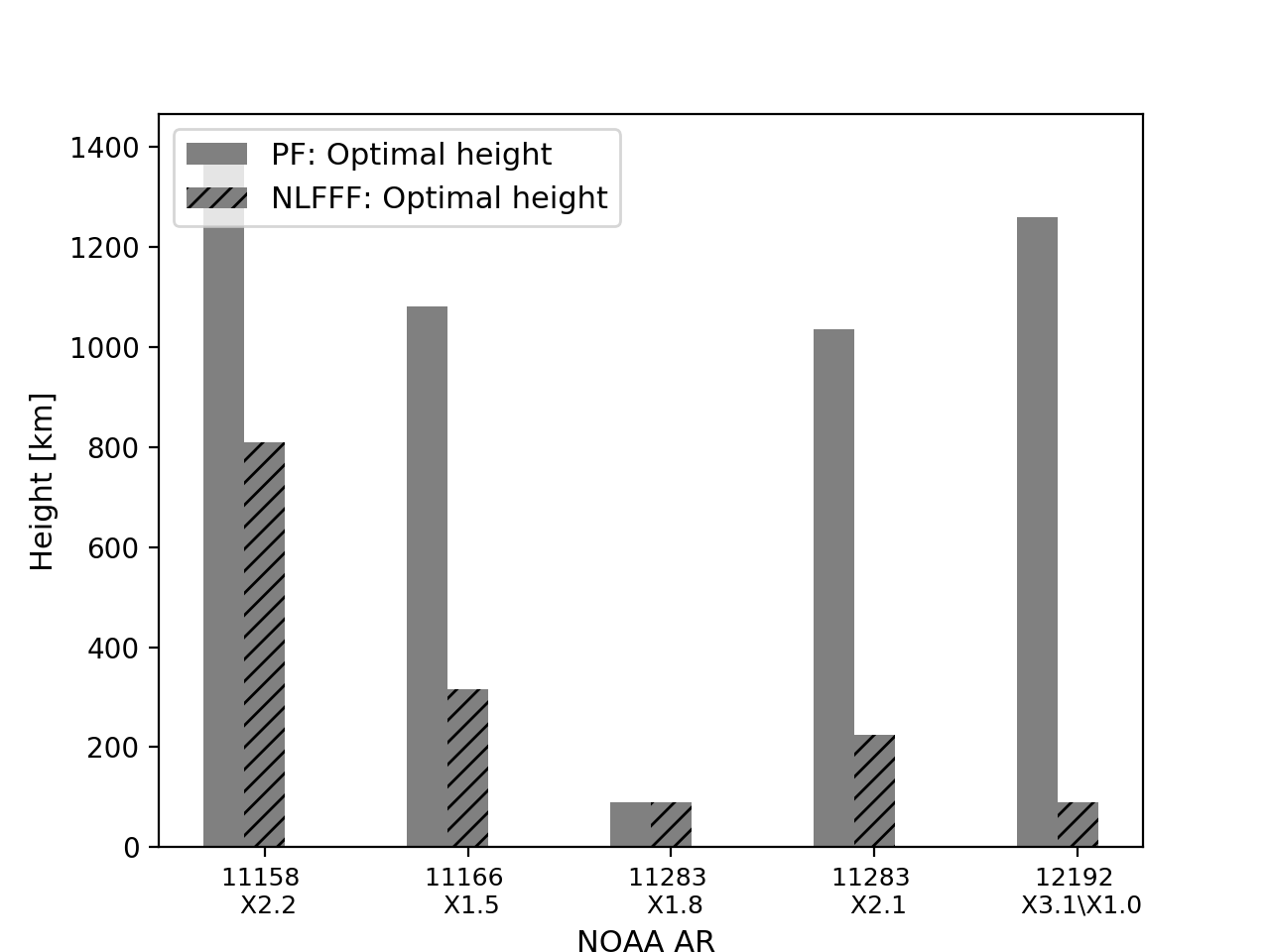}
\put(-240,170){(b)}
\caption{\label{PFvsNLFFF} Summary of $WG_M$ lead times (a) and optimal heights (b) for PF (grey columns) and NLFFF (line-crossed columns) extrapolations in case of four different, eruptive ARs and five major flares. }
\end{figure}

Here, we outline the results of investigating three additional ARs that demonstrate how to further improve the flare onset time prediction capability of the $WG_{M}$ method by means of an optimal height analysis.
For the detailed comparative analysis of ARs 11166, 11283 and AR 12192 see Tables~\ref{1table} in Appendix~\ref{AppendixPFvsNLFFF}. A visual summary of these results is given in Fig.~\ref{PFvsNLFFF} . 

In Fig.~\ref{PFvsNLFFF}a, the columns show the gained time at end of the converging phase ($T_{Imp}^{M}$) at the optimal height. In Fig.~\ref{PFvsNLFFF}b, the columns represent the optimal height of the particular flare events. In Figures~\ref{PFvsNLFFF}a and b, the grey/line-crossed columns refer to PF/NLFFF extrapolations. The plotted values expressed in numbers are given in Table~\ref{1table}. The abscissas of Figures~\ref{PFvsNLFFF}a and b are labelled with the name of the AR that hosted the flare in the same order as the name of the ARs are listed in Table~\ref{1table}.

From Fig.~\ref{PFvsNLFFF} and Table~\ref{1table}, we conclude that the optimal heights and the lead time improvements are not identical for the four studied ARs and five major flare cases. 
We also note that, interestingly, the PF has better $T_{Imp}^{M}$ improvement in four cases out of five.
Based on this finding, one might be tempted to use PF in further studies, for computational efficiency. This might be changed in the future, of course, when computational advances allow for the routine application of NLFFF (or, indeed, even more sophisticated modelling) in much shorter times.

\section{Application of PF to more Active Regions} \label{statistic}

Let us now analyse 13 more flaring AR cases (see Table~\ref{2table}), which all satisfy the selection criteria given in Sec.~\ref{3D}. 

\begin{figure}[]
\centering
\includegraphics[width=0.49\textwidth]{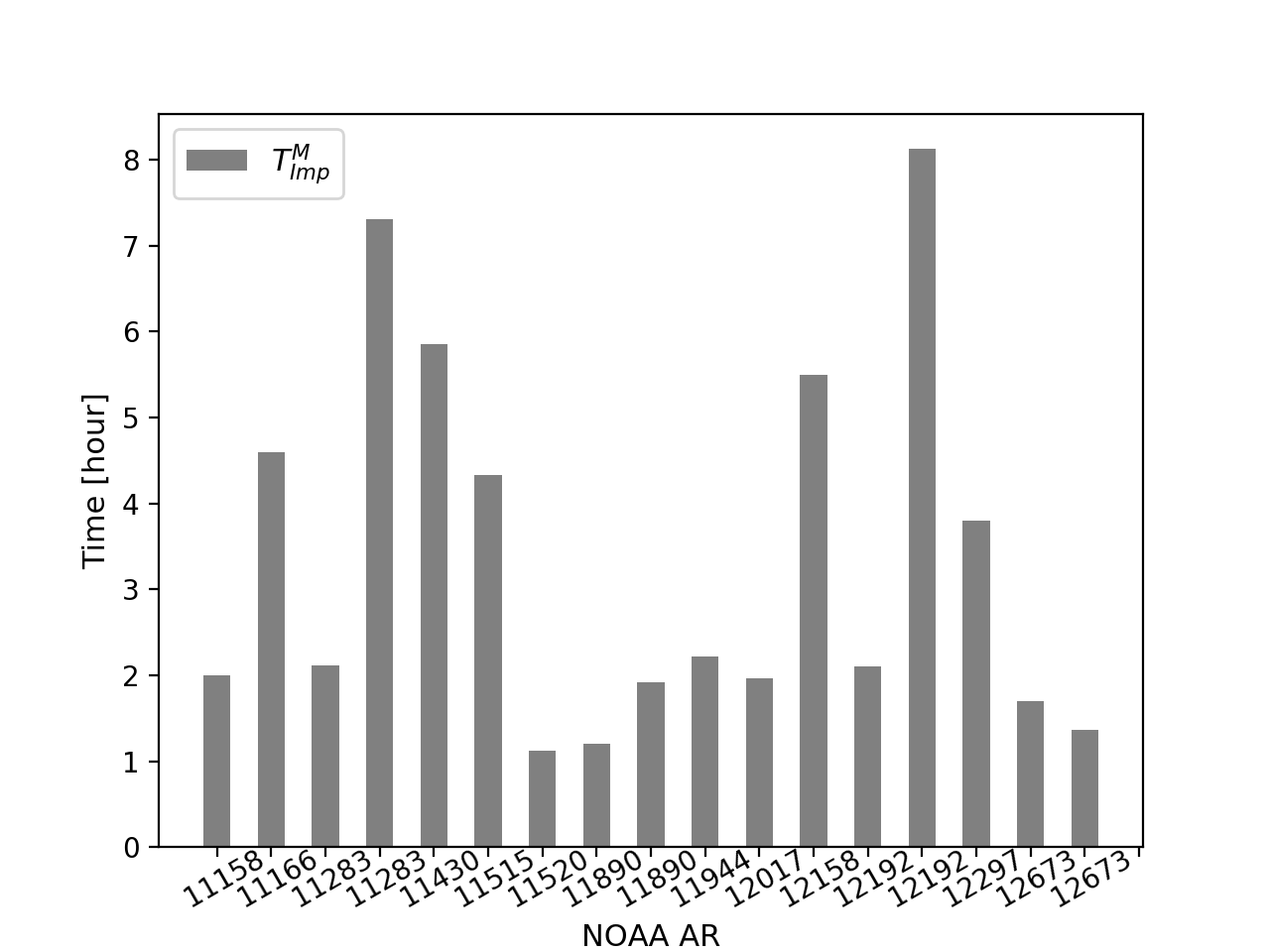}
\put(-240,170){(a)}
\\
\includegraphics[width=0.49\textwidth]{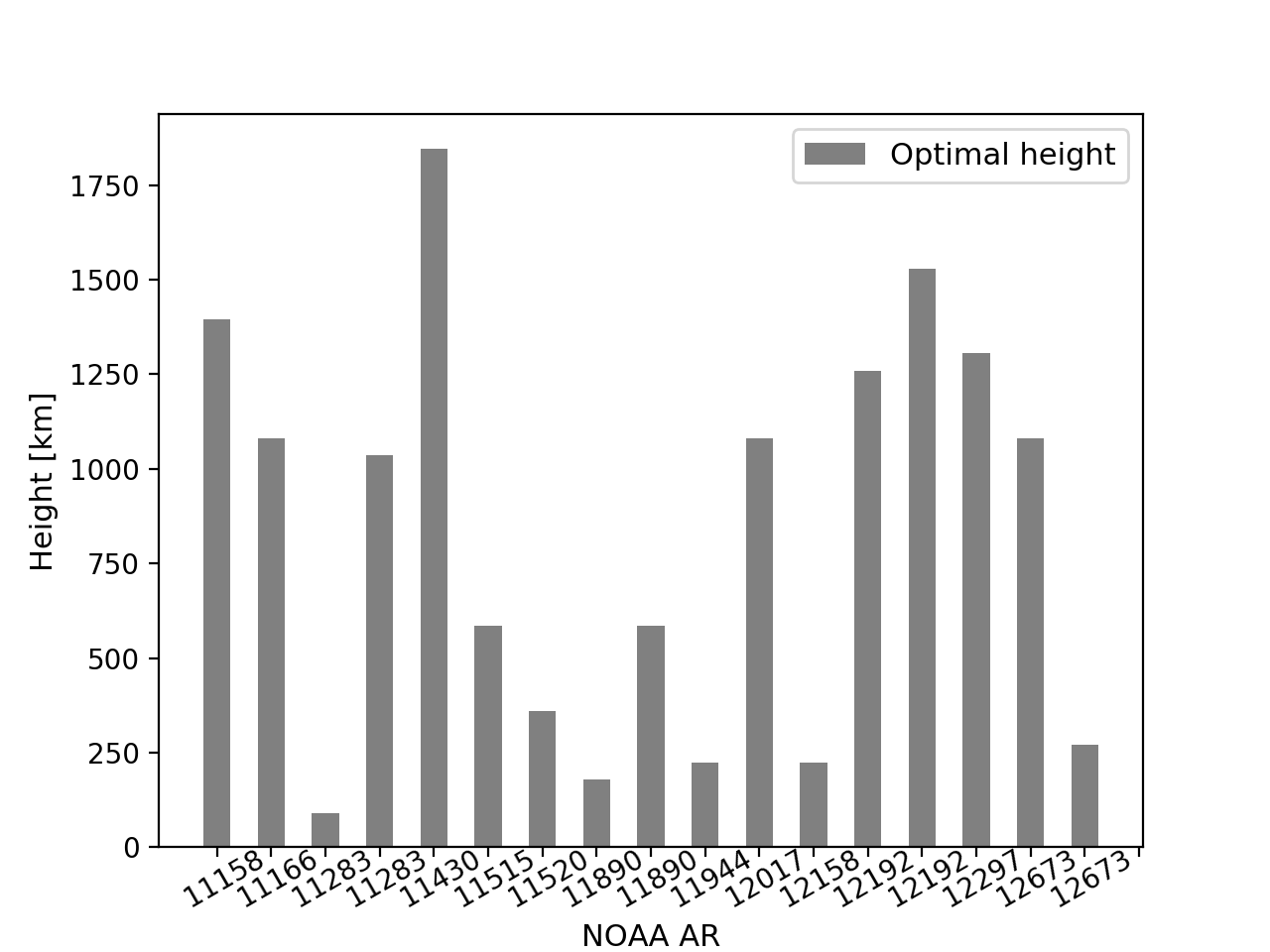}
\put(-240,170){(b)}
\caption{\label{PF} Summary of $WG_M$ lead times (a) and optimal heights (b) for 13 AR cases under the PF extrapolation. For the actual values, see Table~\ref{1table}--\ref{2table}.}
\end{figure}

First, we constructed the 3D PF extrapolations and identified $\delta$-spots. Next, the $WG_{M}$ method was applied to each $\delta$-spot as a function of height with steps of 45 km. The analysis of Section~\ref{casestudy} were carried out to identify the relevant ``inverted V-'' and ``U-shape" of the $WG_{M}$ and $D_{pn}$ parameters. Our findings are summarised in Fig.~\ref{PF} and Table~\ref{2table}.

Similar to the first four examples, we found again that the evolution of the three parameters, i.e.\ $WG_{M}$, $D_{pn}$ and $\Phi$, vary as a function of height in all identified $\delta$-spots. In all cases, the converging phase began earlier and reached its shortest distance also earlier at their respective optimal heights. Furthermore, we note here that we could not identify the concurrent precursor presence of the ``inverted V-'' or ``U-shape" in the cases of non-flaring $\delta$-spots (see two random examples of AR 11158 in Fig. \ref{fig11158f}a and AR 12297 in Fig. \ref{fig12297s}). 

In regards to the 13 flaring ARs studied, we find that (i) the lead time ($T_{Imp}^{M}$) values range between $\sim$ 1 and 8 hours at the identified optimal heights, and (ii) optimal heights seem to fall under two distinct intervals, namely, $\sim$ 90 - 600 km and $\sim$ 1000 - 1800 km.

The estimated flare onset times ($T_{est}$) are much closer to the actual values of $T_{D+F}$ at the optimal height of 1000-1800 km, even considering the 7.2 hr uncertainty. For the 90 - 600 km range, differences can be as large as 2 days between $T_{est}$ and $T_{D+F}$ (see e.g. AR 11515, in Table~\ref{2table}).

Based on our findings above, there is a practically effective optimal height range of 1000-1800 km, for which the prediction capability of the $WG_{M}$ method is considerably improved. In order to determine the potential lead time between 1000 and 1800 km, we use $T_{Imp}^{M}$. This is a very important information because estimating the onset time of a flare seems to rely on the linear relationship between the converging and diverging motions of the opposite polarities \citep[see Fig. 5a of][]{Korsos2019}. Therefore, as a summary we conclude that, based on the values of $T_{Imp}^{M}$ in Tables~\ref{1table}--\ref{2table}, we could estimate the flare onset times $\sim$2-8 hrs earlier with the $WG_{M}$ method at an altitude range of $\sim$1000 - 1800 km above the photosphere using the PF method, rather than working on the photospheric magnetic field.

\section{Conclusions} \label{conclusion}

In this work, we applied the $WG_{M}$ method at different heights between the photosphere and the low corona to 3D magnetic field extrapolations of ARs containing $\delta$-spots at their observed photospheric base, in order to {\it identify an optimal height range where flare prediction could be achieved earlier than using only photospheric data}. 
Our aims were realised by (i) implementing potential (PF) and non-linear magnetic field (NLFFF) exploration techniques and (ii) creating a sample of 3D magnetic maps of sunspots in the lower solar atmosphere. 

As in our previous works, we considered two unique pre-flare patterns of the $WG_{M}$ method (namely, the ``inverted V-shape'' of the $WG_M$ proxy and the ``U-shape'' of the distance ($D_{pn}$) parameter) as a function of height, instead of studying the otherwise popular quantity of the free energy of ARs. 
We still do not have a detailed physical explanation to capture the two pre- flare patterns. However, we put forward our conjecture in Section 4 of  \cite{Korsos2019}. Namely, that a current sheet develops during the convergence phase of the two opposite-polarity area-weighed barycenters, while magnetic reconnection takes place after the end of the divergence phase. The validation of this conjecture should be confirmed by 3D numerical simulations but that is beyond the scope of this work. An alternative suggestion was put forward by \cite{Tlatov2018}, for more details see Fig. 5 of their paper.

In this study, we compared the results obtained by applying the $WG_{M}$ method to PF and NLFFF extrapolation data in four different flaring ARs. We discovered that, at a certain height, called {\it the optimal height}, 
 the fitted ``U-shapes'' enabled us to estimate the expected flare onset time {\it earlier} than using only magnetic data at the photospheric level. This is a key finding of this work.  
 
We also observed that the identified optimal heights and lead-time improvements for estimating the flare onset time vary with the applied extrapolation method. Namely, we found that sometimes the yielded lead-time by PF is better by up to a few hours than using NLFFF extrapolation. This is another important practical aspect because the CPU running time differences between the PF and NLFFF extrapolations are substantial. It might be prohibiting to apply the $WG_{M}$ method under the NLFFF extrapolation in near-realtime due to computational limitations, but the PF extrapolation may offer a viable alternative. 

Next, we restricted to PF extrapolations only. We found that the potential lead-time improvement for estimating the flare onset time varies in the interval (2, 8) hours if we apply the $WG_{M}$ method to an identified ``sunspot" between $\sim$1000 and 1800 km above the photosphere.

In this study, we do not have a negative sample (i.e., non-flaring ARs with $\delta$-spots), because ARs that form $\delta$-spots tend to be flaring \citep[e.g.][and references therein]{Toriumi2019, Georgoulis2019}. It seems that there are $\delta$-spots in some of the ARs studied, here, which do not show the `` inverted V-" and ``U"-shapes though.
In the future, we will also extend this work in at least two directions: i) carry out magnetic field extrapolations and use the $WG_{M}$ method to determine the evolution of the non-flaring and flaring $\delta$-sunspots with flares of lower GOES class (e.g. M-, and even C-classes); and ii) test the findings of this work, as well as the flare precursor capability of the $WG_M$ method, with a larger statistical sample of ARs. We will also investigate, for near-realtime operational purposes, the time needed to determine whether a given $D_{pn}$ is indeed the minimum and how this time relates to the estimated and actual onset time.
\FloatBarrier

\section{ Acknowledgements} 
     MBK is grateful to the University of Sheffield for the supports received while carrying out this research and STFC grant ST/S000518/1 to Aberystwyth University. MBK also acknowledges the open research program of CAS Key Laboratory of Solar Activity, National Astronomical Observatories, No. KLSA201610. MBK and RE acknowledges the CAS Key Laboratory of Solar Activity, National Astronomical Observatories  Commission for Collaborating Research Program for support received to carry out part of this work. RE acknowledges the CAS Presidents International Fellowship Initiative, Grant No. 2019VMA052. RE is also grateful to the Science and Technology Facilities Council (STFC, grant number ST/M000826/1) UK and the Royal Society for enabling this research. SKB acknowledges the support by the National Natural Science Foundation of China (NSFC Grant No. 11750110422, 11433006, 11790301, and 11790305). CJN thanks the Science and Technology Facilities Council (STFC) for the support received to conduct this research through grant number: ST/P000304/1. S.Y. is supported by NSF grants AGS-1654382, AGS-1723436, and AST-1735405 to NJIT. YY acknowledges NSFC supports with the grant No. 11790300 and11790301.

\appendix

\section {Application of the $WG_{M}$-method at different atmospheric heights}
\subsection{AR 11158}\label{AR11158case}

\begin{figure*}[h!] 
\centering
\includegraphics[width=0.39\textwidth]{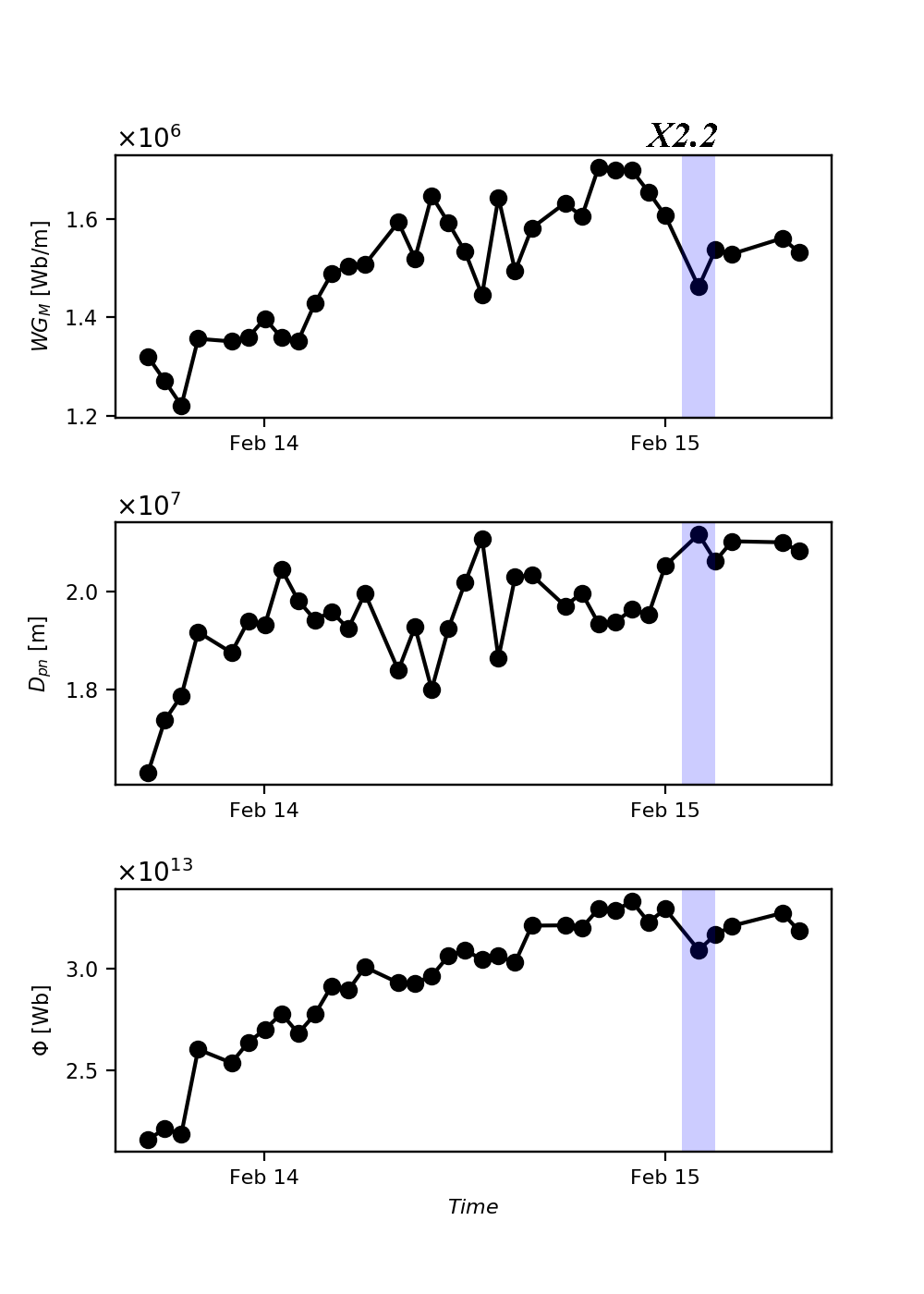}
\put(-210,260){(a)}
\put(-125,260){\normalsize Photosphere}
\includegraphics[width=0.39\textwidth]{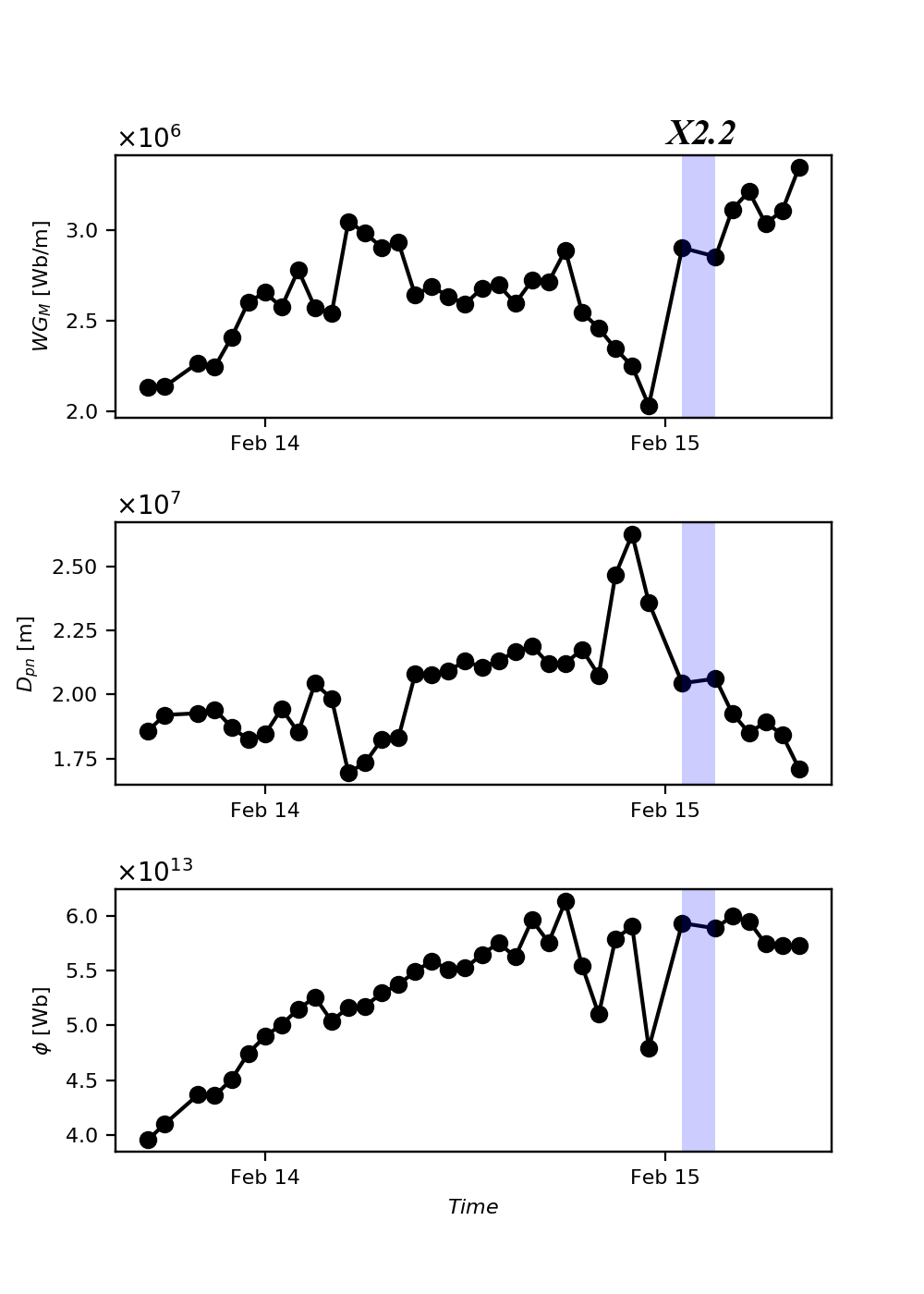}
\put(-210,260){(b)}
\put(-125,260){\normalsize Photosphere}
\caption{\label{fig11158f} Columns (a) and (b) show the graphical visualisation of the result of the $WG_{M}$ analysis for the 1$^{st}$ $\delta$-spot of AR 11158 (for the context image see Fig.~\ref{PIL}a) at the photosphere. Column (a) is the PF and (b) the NLFFF extrapolation case, respectively. The 1$^{st}$ $\delta$-spot was not the cradle of the X2.2 flare.}
\end{figure*}

\begin{figure*}[h!]
\centering
\includegraphics[width=0.39\textwidth]{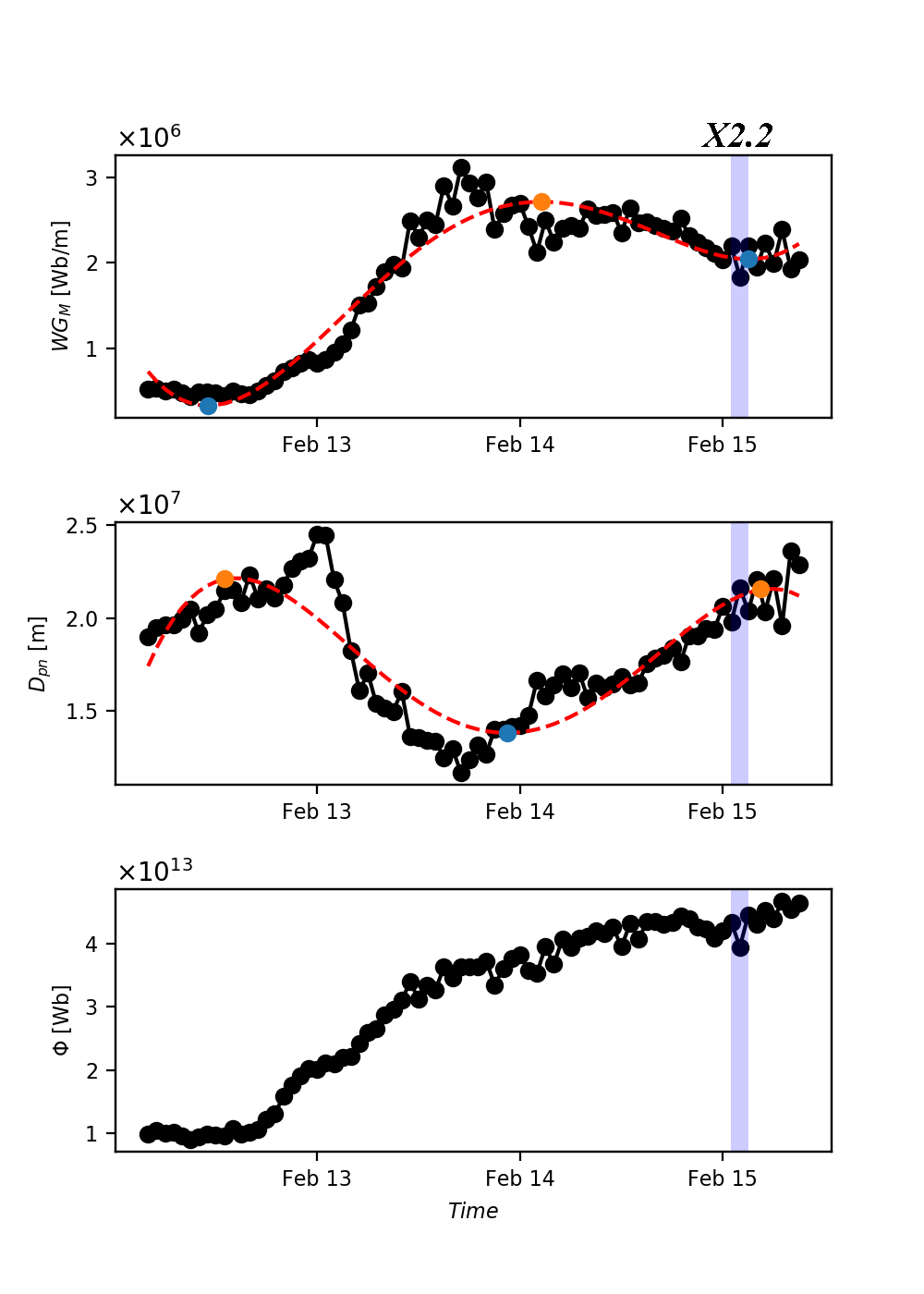}
\put(-210,260){(a)}
\put(-125,260){\normalsize Photosphere}
\includegraphics[width=0.39\textwidth]{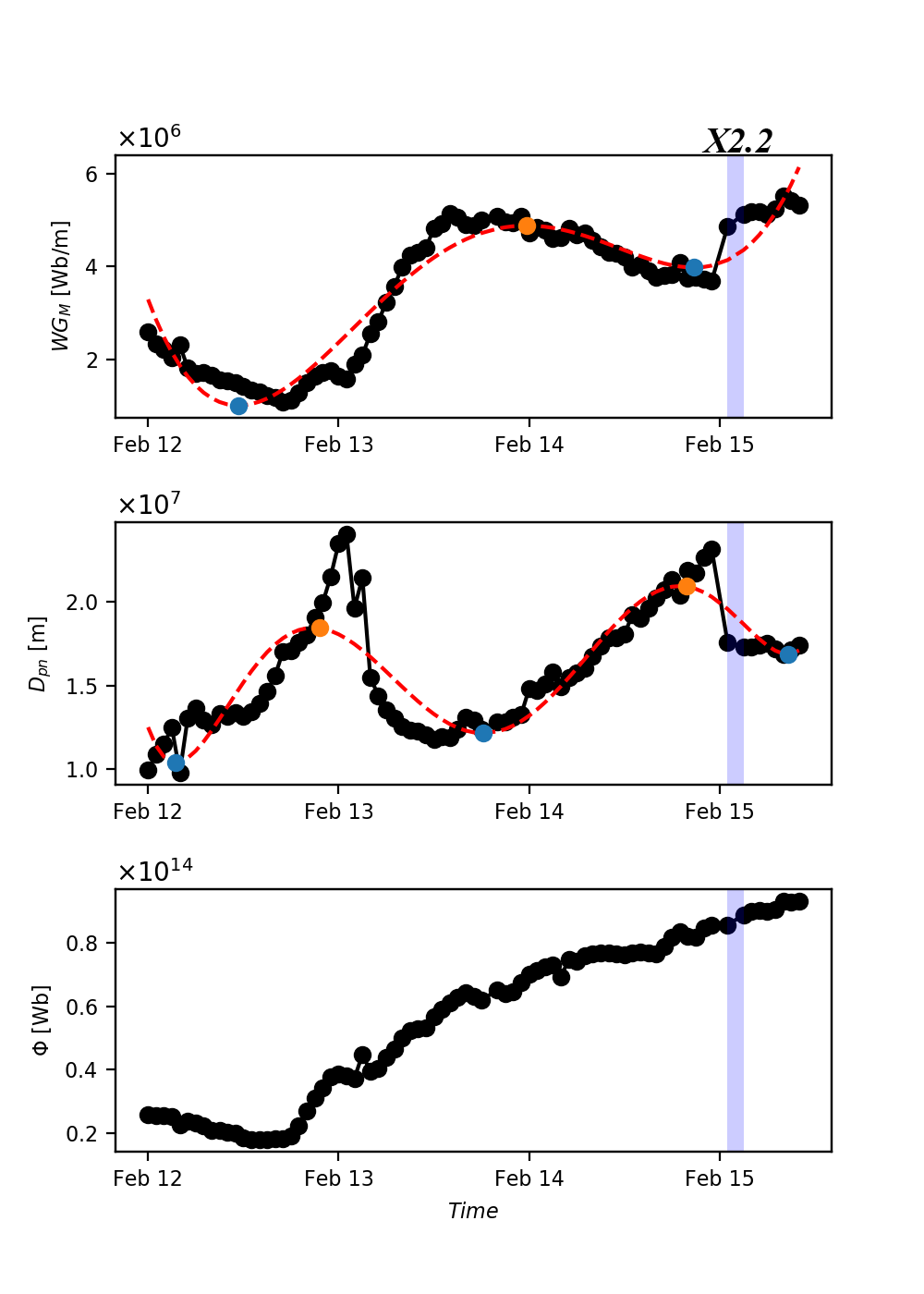}
\put(-210,260){(b)}
\put(-125,260){\normalsize Photosphere}
\caption{\label{fig11158s} Same as Fig.~\ref{fig11158f} but for the 2$^{nd}$ $\delta$-spot of {\it AR 11158} at the photosphere. To identify the ``inverted V-'' and ``U-shape" pre-flare features, we use the maximum (orange dots) and minimum (blue dots) values of the best {\it n}th degree polynomial fit (red dashed line) to the $WG_{M}$ data and to the $D_{pn}$ data. The blue shaded vertical line marks the X2.2 flare occurrence time. This 2$^{nd}$ $\delta$-spot was the host of X2.2 flare. }
\end{figure*}

\begin{figure*}[h!] 
\centering
\includegraphics[width=0.39\textwidth]{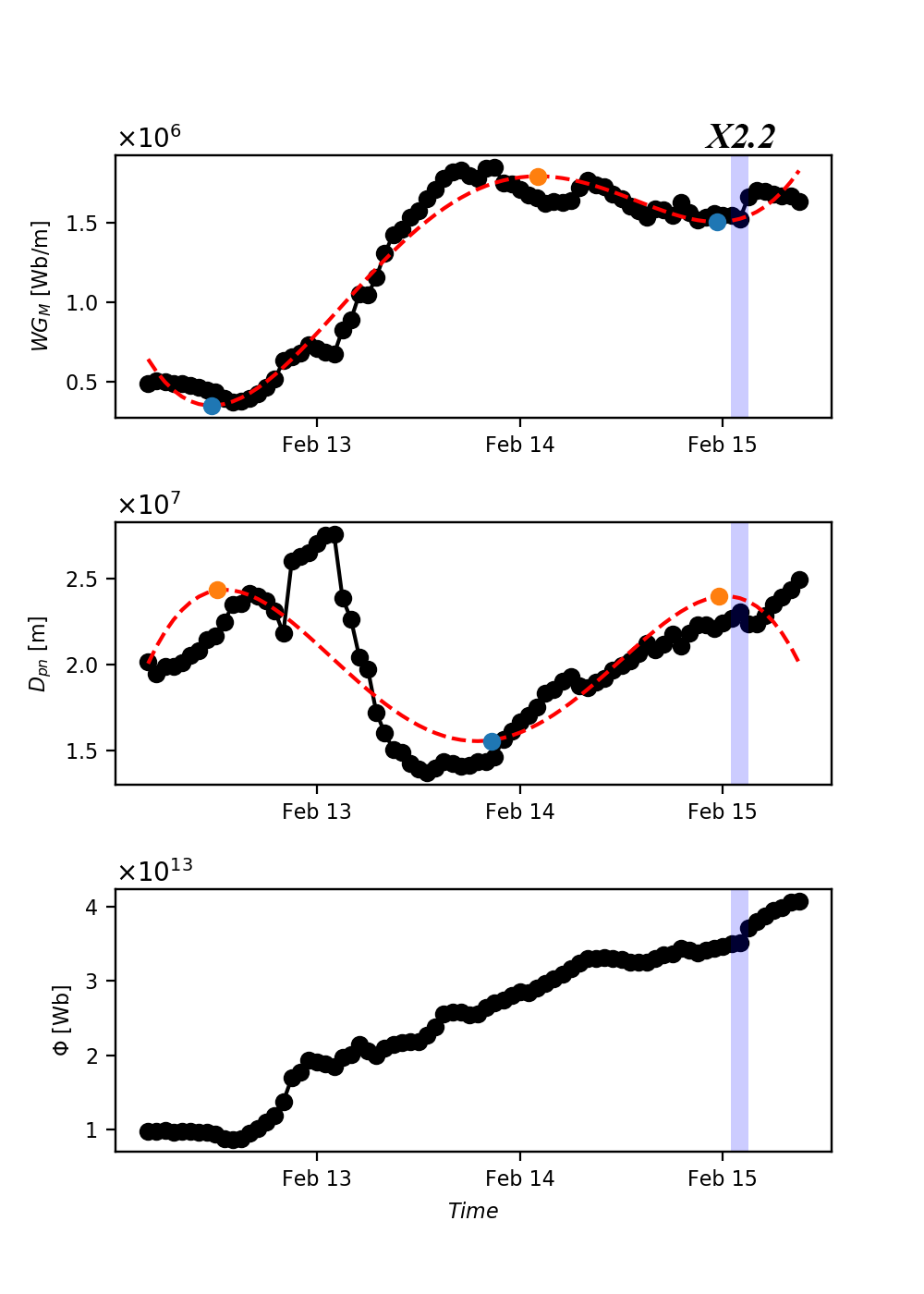}
\put(-210,260){(a)}
\put(-150,260){\normalsize 1395 km above Photosphere}
\includegraphics[width=0.39\textwidth]{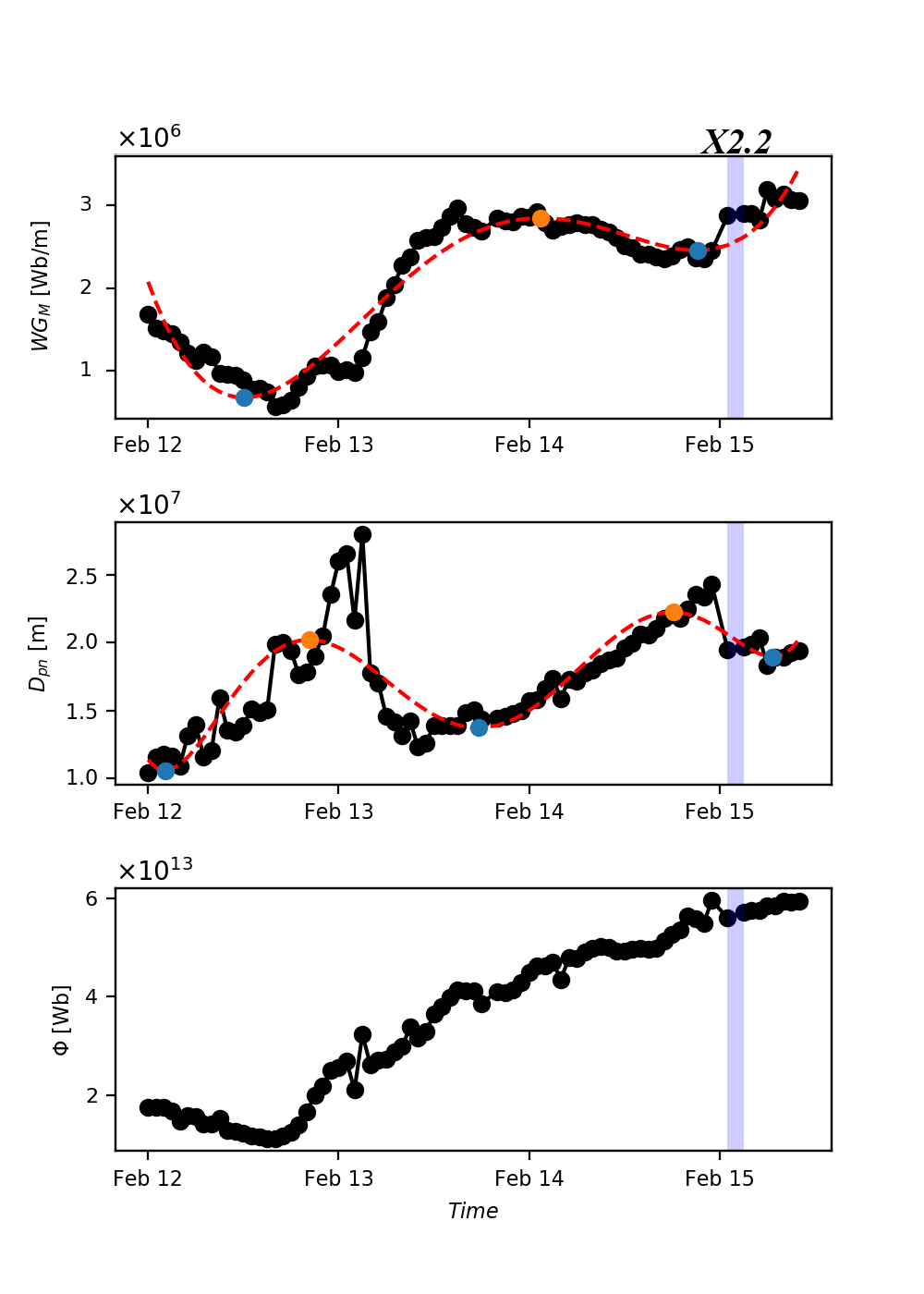}
\put(-210,260){(b)}
\put(-150,260){\normalsize 810 km above Photosphere}

\caption{\label{fig11158t} Same as Fig.~\ref{fig11158s} for {\it AR 11158} where panels (a) and (b) now illustrate the evolution of the flare precursor parameters at the {\it optimal height}. (a): The optimal height is 1395 km above the photosphere in the PF case. (b): The optimal height is 810 km from the photosphere in the NLFFF case.}
\end{figure*}

Figures \ref{fig11158f}-\ref{fig11158t} show the evolution of $WG_{M}$, $D_{pn}$ and $\Phi$ before the X2.2 flare, which occurred at 01:56 on 15/02/2011 in AR 11158. Panels (a) and (b) of each figure reveal the evolution of the various pre-flare indicators, applied to the PF and NLFFF extrapolations. The upper panels in each figure are the temporal variations of $WG_{M}$. The pre-flare behavior of $WG_{M}$ is fitted by an {\it nt}h-order polynomial (red line), where the orange dot corresponds to the maximum of $WG_{M}$. The middle panels demonstrate the evolution of $D_{pn}$. The consecutive maximum-minimum-maximum (orange-blue-orange dots) locations of the fitted {\it nt}h-degree polynomial denote the full converging-diverging phase uncovered by $D_{pn}$. The vertical blue stripes mark the flare peak time. The bottom panels show the evolution of the unsigned magnetic flux ($\Phi$). 

To find the best {\it nt}h-order polynomial, we fit the data with a range of polynomial degrees and pick the degree that has the lowest root mean square error. 

Figures \ref{fig11158f}-\ref{fig11158s} show the evolution of $WG_{M}$, $D_{pn}$ and $\Phi$ at the photospheric level in the case of 1$^{st}$ and 2$^{nd}$ $\delta$-spots of AR 11158. Furthermore, Fig.~\ref{fig11158t} (a)-(b) correspond to findings obtained at the identified optimum heights in the case of PF an NLFFF analyses of 2$^{nd}$ $\delta$-spot, respectively.

\FloatBarrier

\FloatBarrier

\FloatBarrier

\subsection{AR 12297} \label{negative}

\begin{figure}[h!]
\centering
\includegraphics[width=0.45\textwidth]{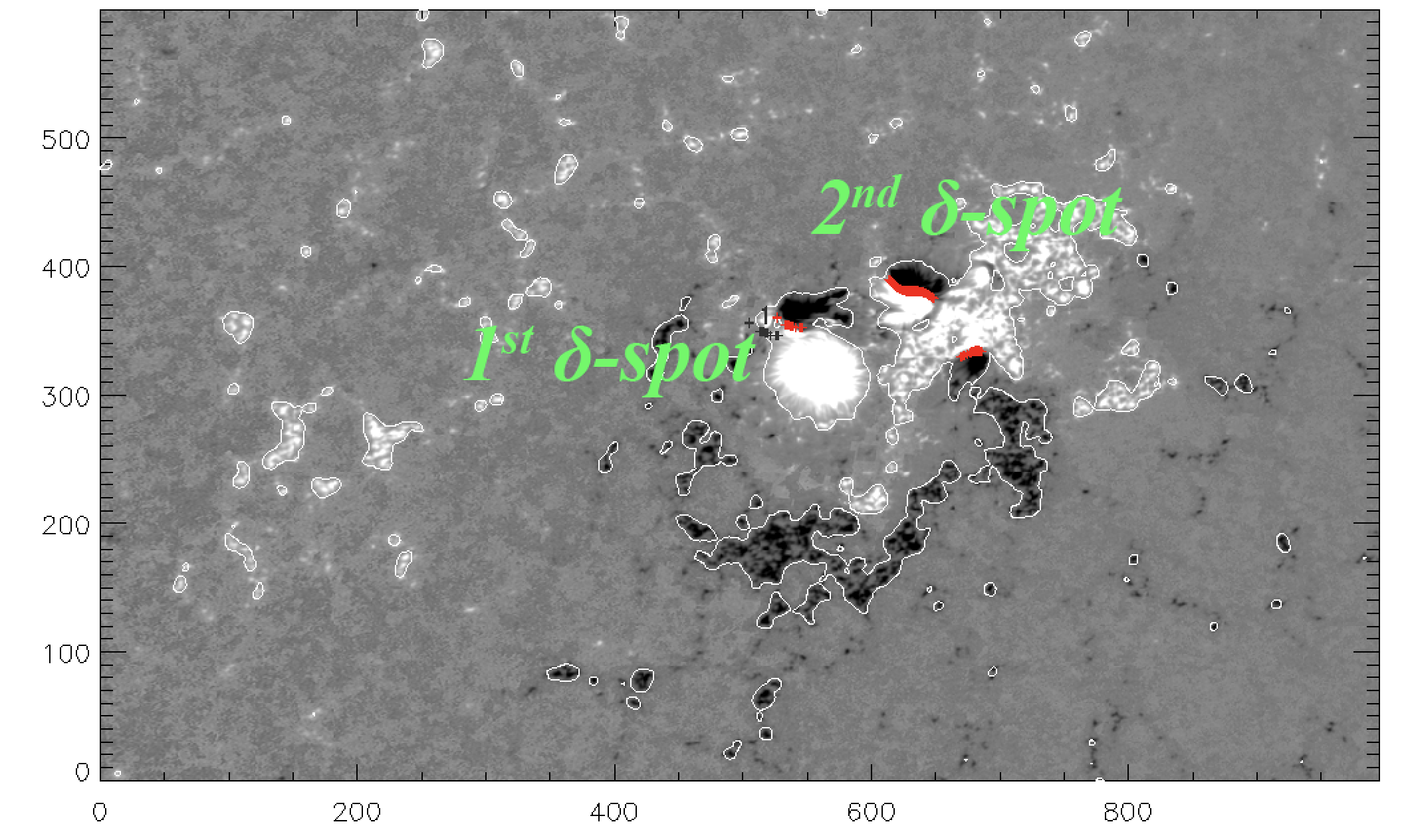}
\caption{\label{12297PIL} Magnetogram snapshots showing the analysed two $\delta$-spots of {\it AR 12297} on 10/03/2015 at 22:00:00. The red dotted lines are the automatically identified PILs of the AR. The white countered areas show the identification of strong flux elements.}
\end{figure}

\begin{figure*}[h!]
\centering
\includegraphics[width=0.39\textwidth]{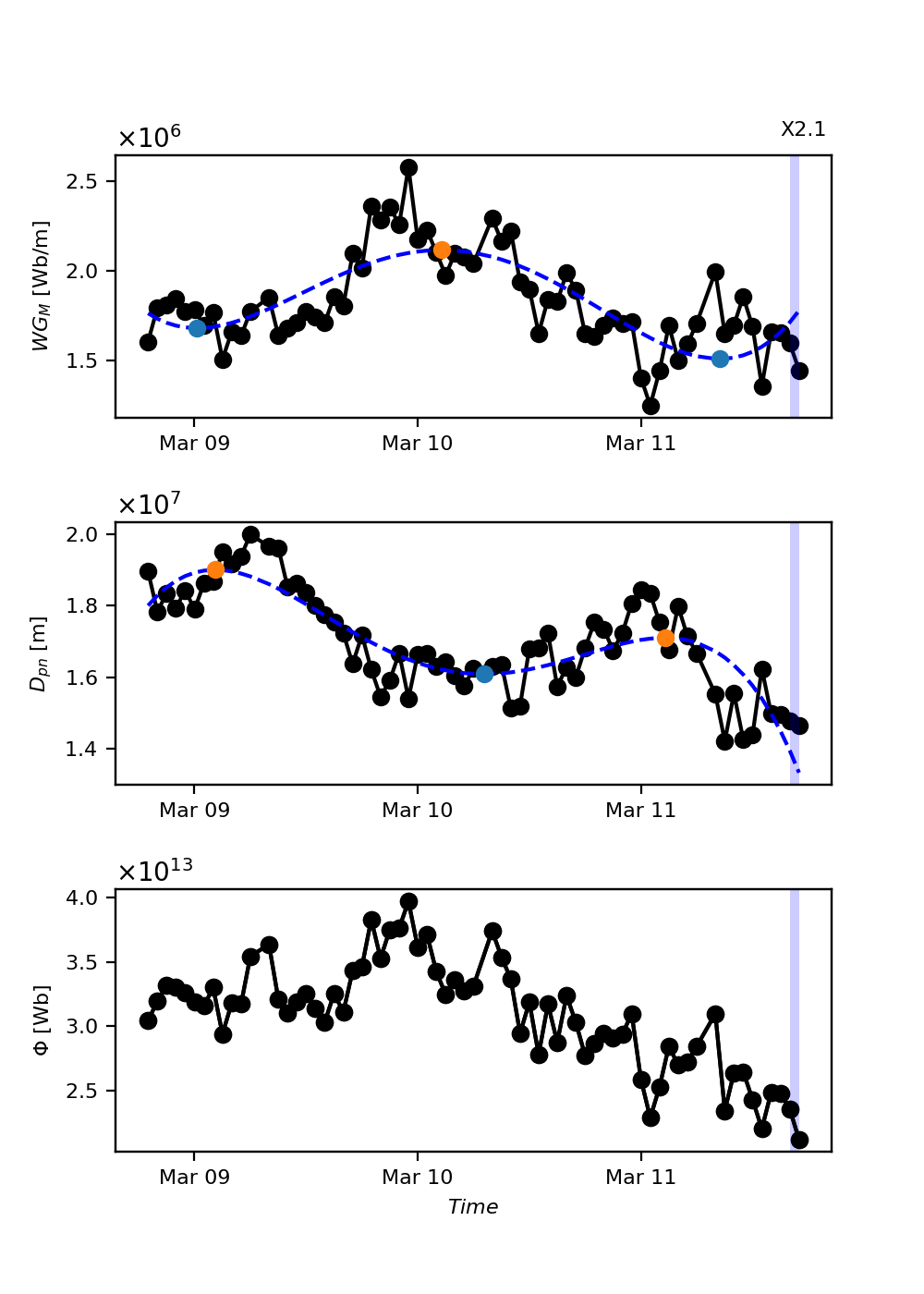}
\put(-210,260){(a)}
\put(-125,260){\normalsize Photosphere}
\includegraphics[width=0.39\textwidth]{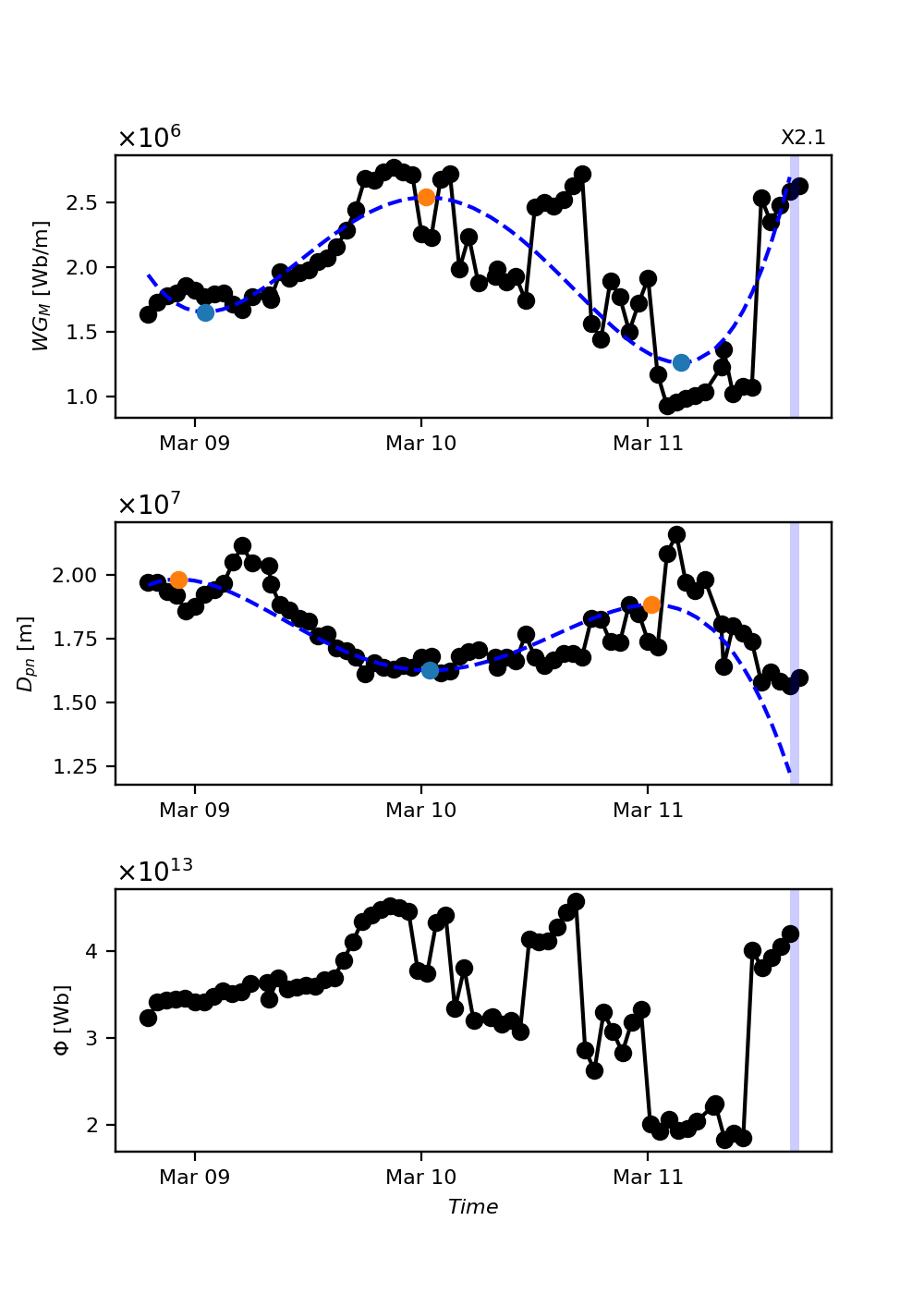}
\put(-210,260){(b)}
\put(-150,260){\normalsize 500 km above Photosphere}
\caption{\label{fig12297f} Columns (a) and (b) show the graphical visualisation of the result of the $WG_{M}$ analysis for the 1$^{st}$ $\delta$-spot of {\it AR 12297} (for the context image see Fig.~\ref{12297PIL}). Column (a) is for the photosphere and panel (b) the 500 km level case, respectively.  }
\end{figure*}

\begin{figure*}[h!]
\centering
\includegraphics[width=0.39\textwidth]{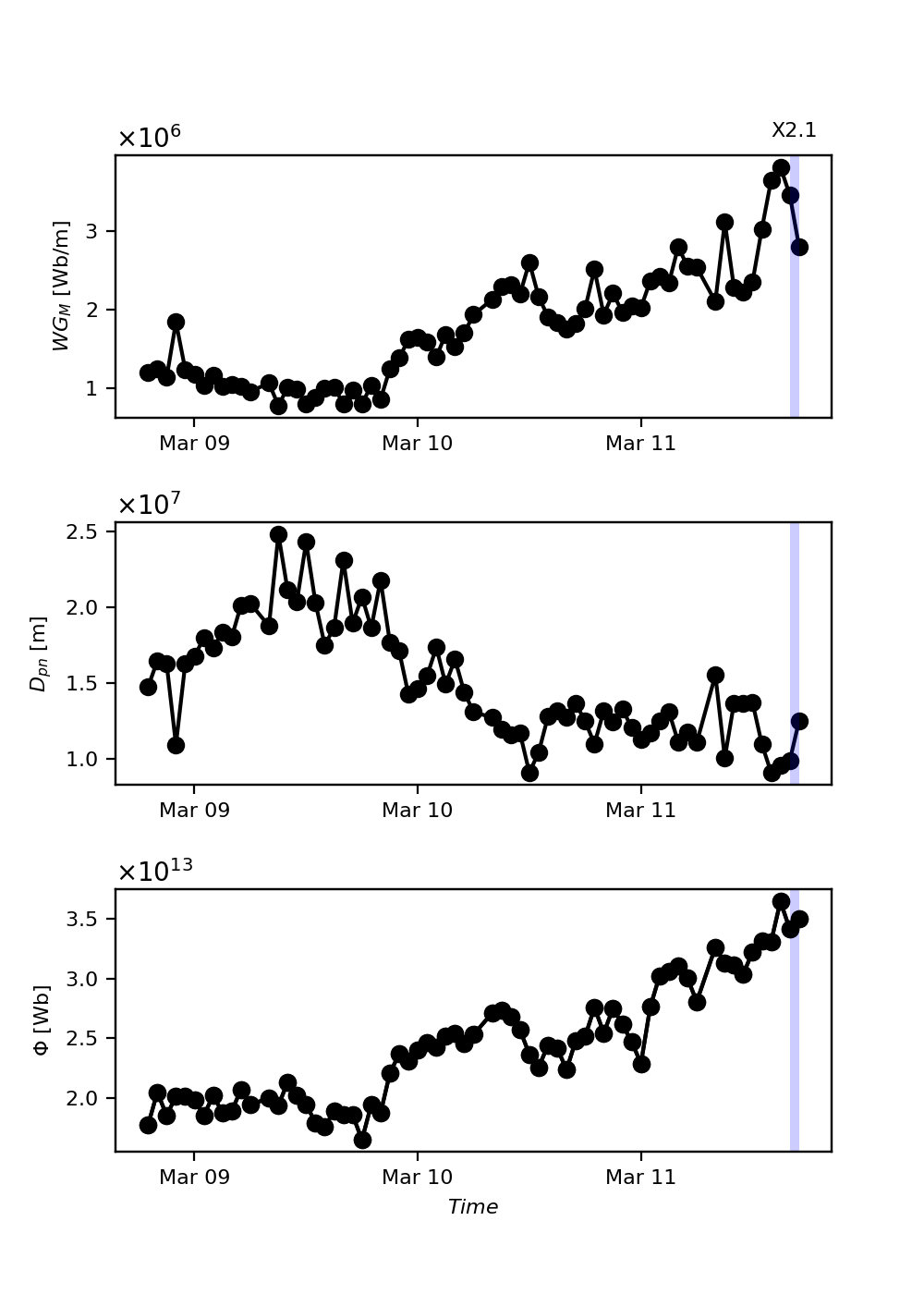}
\put(-210,260){(a)}
\put(-125,260){\normalsize Photosphere}
\includegraphics[width=0.39\textwidth]{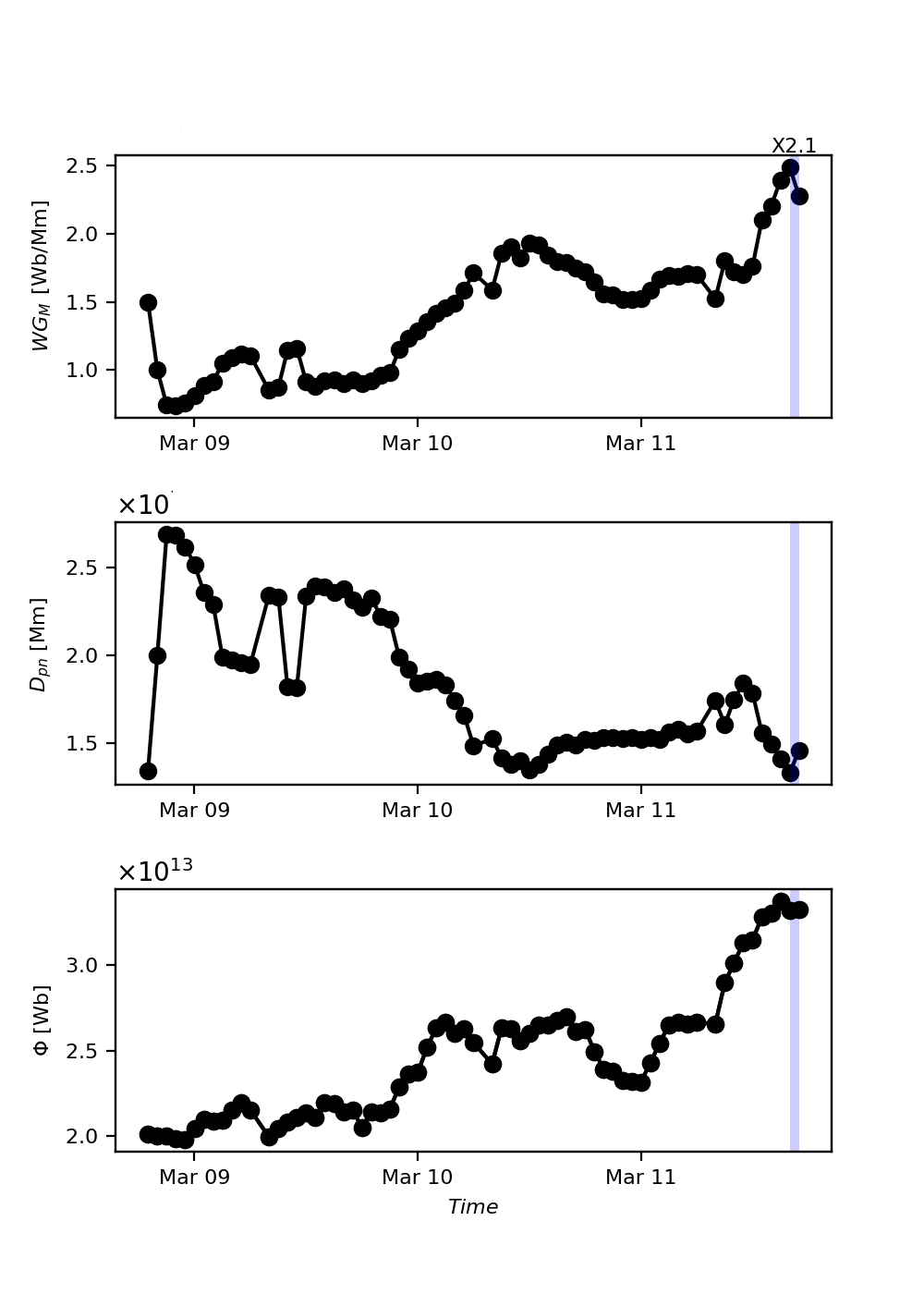}
\put(-210,260){(b)}
\put(-150,260){\normalsize 500 km above Photosphere}

\caption{\label{fig12297s} Same as Fig.~\ref{fig12297f} but for the 2$^{nd}$ $\delta$-spot of {\it AR 12297}. }
\end{figure*}

The AR 12297 was host of an X2.1 flare at 16:22 on 11/03/2015. This AR has two $\delta$-spots determined by the method of \cite{Cui2006}, labelled as the 1$^{st}$ and 2$^{nd}$ $\delta$-spots, respectively, in Fig.~\ref{12297PIL}.
Figures \ref{fig12297f}-\ref{fig12297s} show the temporal variation of $WG_{M}$, $D_{pn}$ and $\Phi$ at the photosphere (Panel a); and 500 km above the solar surface (Panel b), applied to the PF magnetic extrapolation data. Panels (a) and (b) of Figures \ref{fig12297f} reveal the evolution of ``inverted V-'' and ``U-shape" of the $WG_{M}$ and $D_{pn}$ parameters before X2.1 flare, in case of the 1$^{st}$ $\delta$-spot. Note, in Fig. \ref{fig12297s}, we cannot identify the ``inverted V-'' and ``U-shape"  before the X2.1  in the case of 2$^{nd}$ $\delta$-spot. The reason is because actually the 1$^{st}$ $\delta$-spot was the source of the X2.1 flare \citep{Lu2019}. 

\section{Brief comparison analyses of pre-flare behaviour of further three Active regions based on PF and NLFFF extrapolations} \label{AppendixPFvsNLFFF}
\subsection{AR 11166} \label{11166}

\begin{figure}[h!]
\centering
\includegraphics[width=0.55\textwidth]{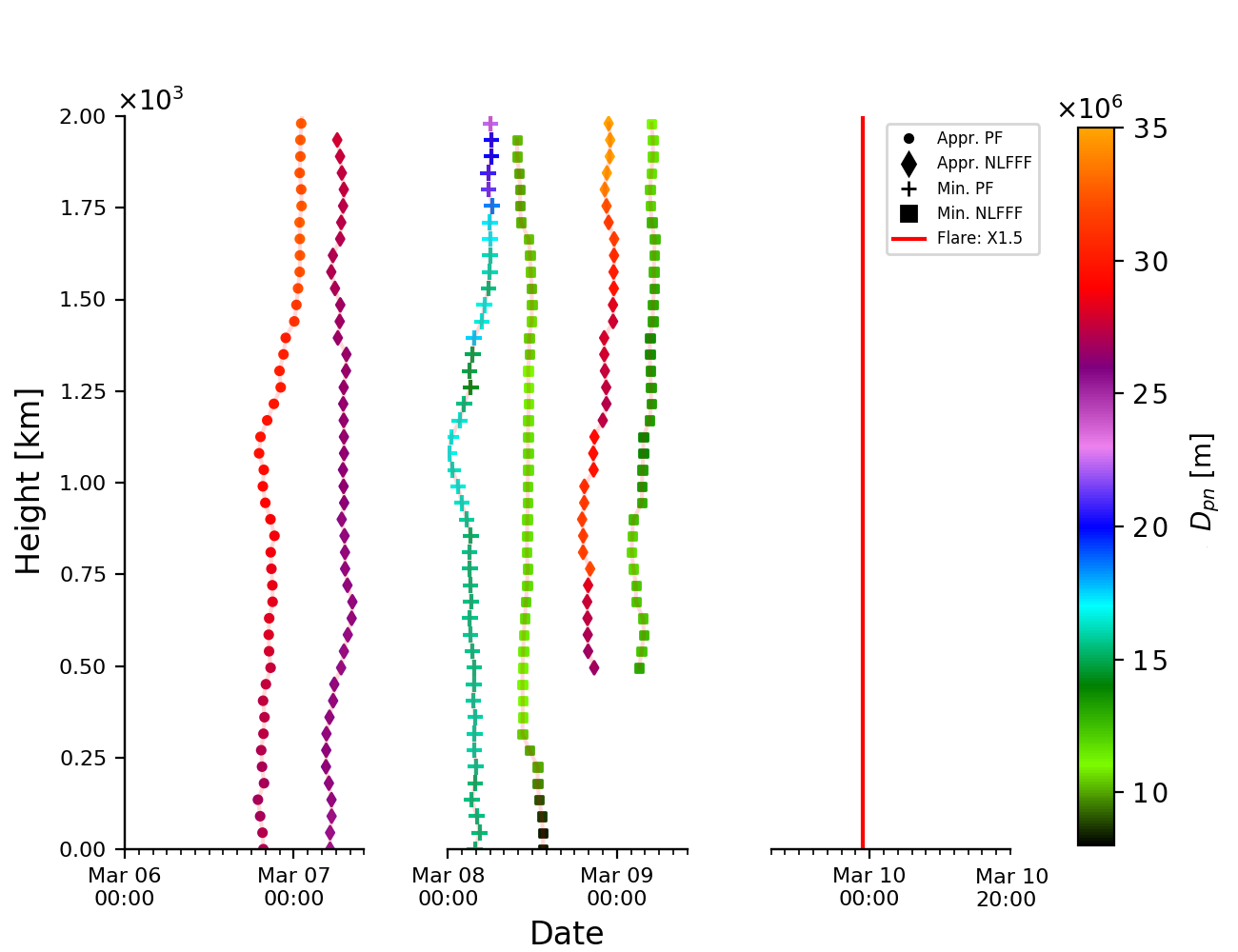}
\put(-142,200){\scriptsize 2$^{nd}$}
\put(-153,193){\scriptsize Appr. Min.}
\put(-170,200){\scriptsize 1$^{st}$}
\put(-170,193){\scriptsize Min.}
\put(-210,200){\scriptsize 1$^{st}$ }
\put(-210,193){\scriptsize Appr.}
\caption{\label{fig11166t} Same as Fig.~\ref{11158height} but for {\it AR 11166}.}
\end{figure}

The second example is {\it AR 11166}. Here, we investigate the pre-flare states before an X1.5 flare. This flare occurred in the single $\delta$-spot of the AR at 23:23 on 09/03/2011 \citep{Vemareddy2014}. 
We could identify the prominent and typical ``inverted V'' and ``U" shapes prior to X1.5 in the vertical region from the photosphere up to 2000 km at each 45 km step.  
In the NLFFF extrapolation case, we further noticed that two consecutive precursors of the $WG_{M}$ and the $D_{pn}$ appear instead of one only above 500 km. 
Therefore, we can uniquely identify the optimal height only in the case of the first modelling approach. In the second case, we do not have photospheric reference data (see Fig.~\ref{fig11166t}).
In the PF case, the optimal height is identified at 1080 km where the converging phase started ($T_{Imp}^{C}$) 0.6 hrs earlier and ended ($T_{Imp}^{M}$)  4.6 hrs earlier than in the photosphere. In the NLFFF case, $T_{Imp}^{C}$ is 0.5 hrs and $T_{Imp}^{M}$ is 2.9 hrs corresponding to an optimal height of 315 km. 
Here, we could estimate the flare onset time a couple of hours earlier using either of the extrapolations. Unfortunately, $T_{est}$ seems to be rather overestimated, with 8 hrs, in the PF extrapolation case. However, the $T_{est}$ value is well in agreement with the $T_{D+F}$ value when applying data from the corresponding NLFFF extrapolation.
In the PF case, using Eq. of Fig. 5 of \cite{Korsos2019}, we estimated the $S_{flare}$ as an M-class that is an underestimate when compared to the measured X1.5 flare intensity. In the NLFFF case, $S_{flare}$ is found to be correctly. 
\FloatBarrier

\FloatBarrier

\subsection{AR 11283} \label{11283}

The next example is {\it AR 11283} with a flare of X1.8 that occurred at 22:20 on 06/09/2011 and with another one of X2.1 at 22:38 on 07/09/2011. These two flares had their cradle in the same $\delta$-spot of the AR \citep{Liu2014}. Here, the characteristic pre-flare behavior of the $WG_{M}$ and the $D_{pn}$ are evaluated, using the appropriate 3D constructed magnetic field structures, where both precursor patterns are identifiable prior to each of the two flares. We found that the two flare precursor behaviours of the X1.8 flare disappear from 1000 km upwards in both of the PF and NLFFF extrapolation modelling. 
For the X1.8 flare, the optimal height of the PF is found to be at 90 km and for the NLFFF it is also at 90 km, where, the value of $T_{Imp}^{C}$ is 1.6/1.2 hrs and the value of $T_{Imp}^{M}$ is 2.1/1.5 hrs in the PF/NLFFF case, respectively.

\begin{figure}[h!]
\centering
\includegraphics[width=0.53\textwidth]{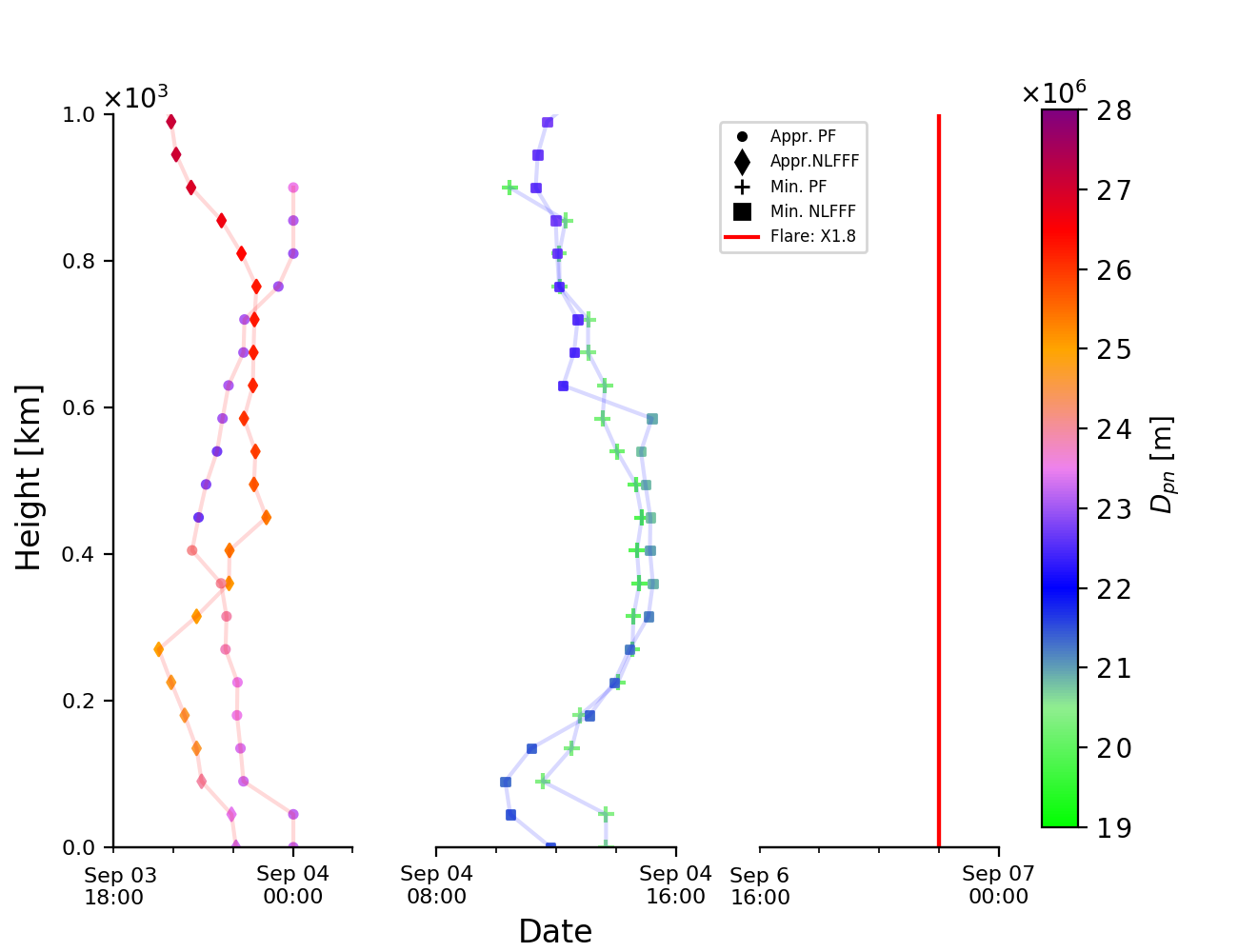}
\put(-270,210){(a)}
\\
\includegraphics[width=0.53\textwidth]{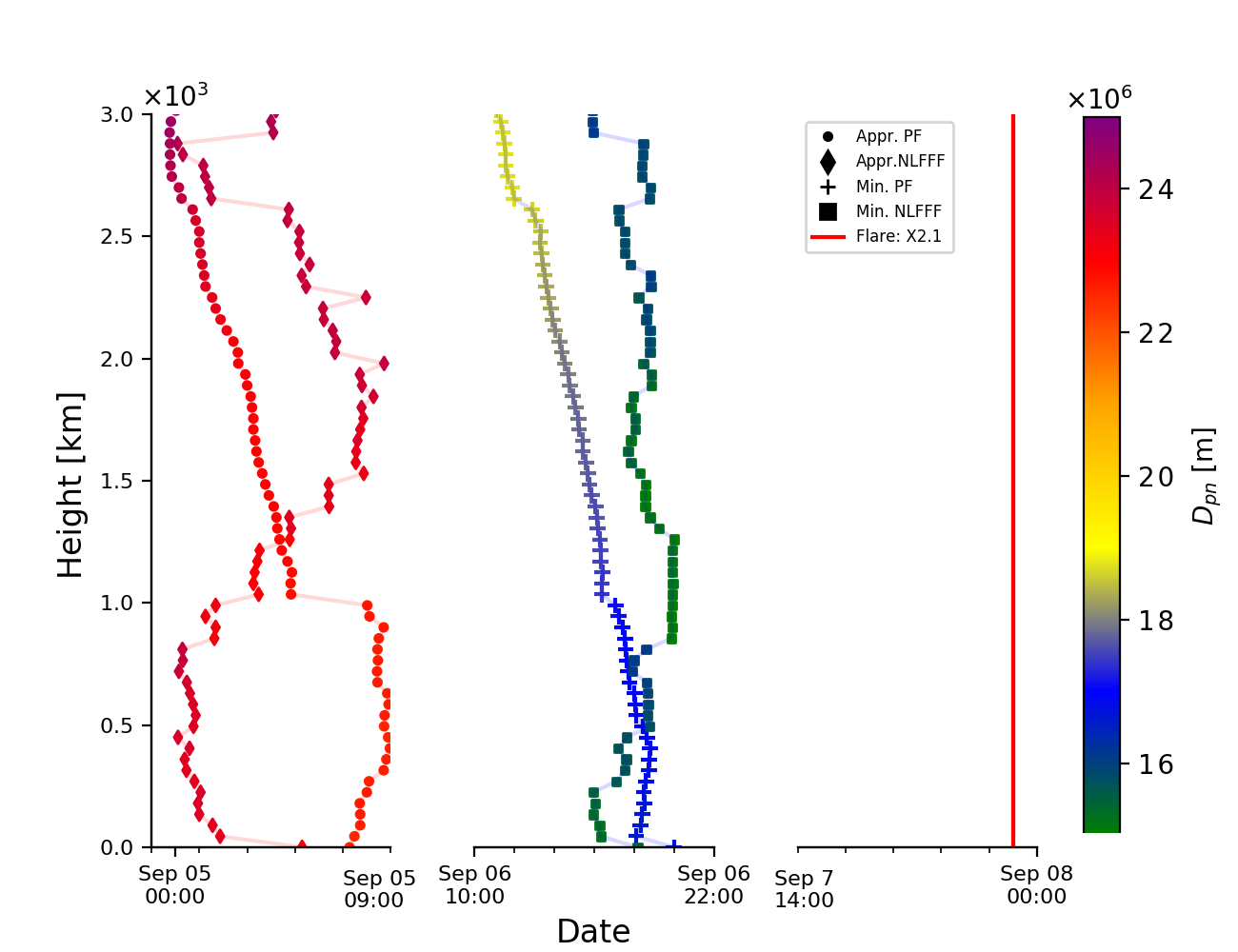}
\put(-270,210){(b)}
\caption{\label{11283height} The plots (a) and (b) correspond two X-class flares (X1.8 and X2.1 flares) of {\it AR 11283}.}
\end{figure}

For the X2.1 flare, the optimal height is found to be 1035 km for PF magnetic field structures of AR 11283. The converging phase begun 6.7 hours beforehand and finished 7.3 hrs earlier at the optimum height 1035 km when compared to the result of analysis applied to the data in the photosphere.  The 1035 km was chosen for the PF as the optimum height because the beginning and finishing moments of the converging phase started to shift  continuously above this height until 3000 km (see Fig.~\ref{11283height} b). Extrapolation was carried out only up to 3000 km.  In the NLFFF case, the optimum height is found to be at 225 km, where, the value of $T_{Imp}^{C}$ is 3.8 hrs and the value of $T_{Imp}^{M}$ is 1.8 hrs, respectively.

In summary, the overall situation with the estimates is similar to that of the {\it AR 11166} in the PF case. Here, in both extrapolation models, the $T_{est}$ values are underestimated for the X1.8 flare while $T_{est}$ is overestimated for the X2.1 case, just like in the PF case for AR 11166. $S_{flare}$ are fairly well estimated. In the case of AR 11283, we could estimate the onset time of the X1.8 flare 5.5 hrs earlier with the PF data when compared to the counterpart obtained with the NLFFF extrapolation. For predicting the X2.1 flare, the PF and NLFFF extrapolations seem to be similarly beneficial.

\FloatBarrier

\FloatBarrier

\subsection{AR 12192} \label{12192}

The last randomly selected example to demonstrate the prediction capability of the $WG_M$ method in a 3D lower solar atmosphere model is {\it AR 12192}, with a series of flares that occurred in the same $\delta$-spot, according to \cite{Bamba2017}. The first pair of characteristic pre-flare behaviors of the $WG_{M}$ and $D_{pn}$ are observed prior to the X1.6 flare, which occured at 14:28 on 22/10/2014. However, the ``U" shape starts to form when the AR is on $\sim$$67^{\circ}$, where the magnetic projection effects are not neglectable. Therefore, here, we did not investigate the case of X1.6 flare.

 Later, another pair of the ``inverted V'' and ``U" shapes are found and evaluated prior to the X3.1 (at 21:41 on 24/10/2014) and X1.0 (at 18:08 on 25/10/2014) flares, respectively. In the cases of the X3.1 and X1.0 flares, the optimal height of the PF approach is at 1260 km, and, it is at 90 km for the NLFFF extrapolation. The lead times are very similar in the case of NLFFF ($T_{Imp}^{M}$  =2.3 hrs) and PF approach ($T_{Imp}^{M}$ =2.1 hrs).  Furthermore, in the case of NLFFF, the $T_{est}$ values is well estimated for the first flare occurrence. The difference between the estimated and the actual occurrence times are close to the $\pm$ 7.2 hrs uncertainty. We cannot say this about the case of PF extrapolation.  
Here, we also must mention that {\it AR 12192} produced a further X2.0 flare at 10:56 26/10/2014. However, we can only observe the two typical pre-flare patterns when using the PF data. 
\FloatBarrier

\begin{figure}[h!]
\centering
\includegraphics[width=0.53\textwidth]{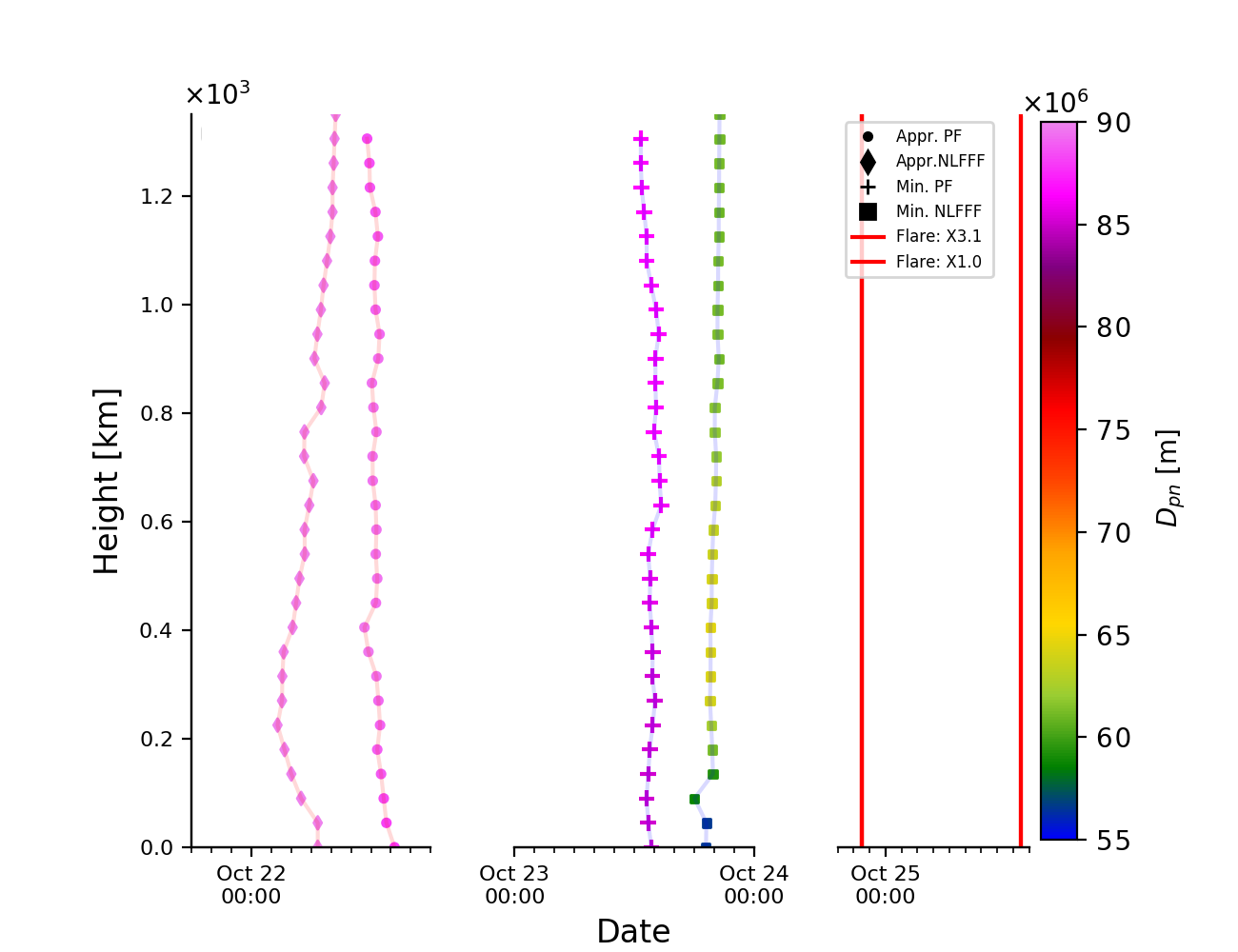}

\caption{\label{12192height} The plot corresponds to two X-class flares (X3.1 and X1.0 flares) of {\it AR 12192}.}
\end{figure}

\FloatBarrier

\begin{table*}[h!] 
\resizebox{\textwidth}{!}{
\begin{tabular}{|c|c|cccccc|cccccc|}
\cline{3-14}
\multicolumn{1}{c}{}  & \multicolumn{1}{c|}{} & \multicolumn{6}{c|}{PF} & \multicolumn{6}{c|}{NLFFF} \\
\hline                   
NOAA   AR  & Flare Intensity &  $T_{Imp}^{C}$  [h]        &  $T_{Imp}^{M}$ [h]           & Opt. Height  [km]    &   $S_{flare}$   &   $T_{est}$ [h]& $T_{D+F}$ [h] &$T_{Imp}^{C}$  [h]          &  $T_{Imp}^{M}$ [h]          & Opt. Height [km] &   $S_{flare}$   &  $T_{est}$ [h]& $T_{D+F}$ [h] \\
\hline 
\hline    

11158	&X2.2	&	1.2	&	2.0&	1395	&X &29.5&42&	0.9	&	0.7	&	810 &X	&30&32.5	\\
\hline 
11166	&	X1.5	&	0.6&	4.6	&	1080	&$>$M5& 39.1	&46.8& 	0.5	&	2.9	&	315&X& 39.5	&36.2	\\
\hline 
\multirow{2}{*}{11283}	&	X1.8	&	1.6&	2.1 	&	90	&X& 22.2	&53.6&	1.2	& 1.5&	90	&X& 21.8	&59.6\\
	&	X2.1	&	6.7	&	7.3	&	1035		&X& 47.4	&33.7&3.8	&	1.8	&	225&X& 50.1&30.1\\
\hline
12192	&	X3.1/X1.0	&	2.5	&	2.1	&	1260	&X&	21.9&43.7/63.7&1.7	&2.3	&90&X& 41.3&32.9/52.9\\
					
\hline

\end{tabular}}

\caption{Table to compare the results of applying the $WG_M$ method, obtained by means of PF and NLFFF extrapolations in four investigated ARs. The table includes of how many hours earlier the converging phase ($T_{Imp}^{C}$) began and reached the minimum value ($T_{Imp}^{M}$) at the optimal height (Opt. Height) when compared to the photosphere. $S_{flare}$  is the estimated flare class. $T_{est}$ is the estimated flare onset time in hrs. $T_{D+F}$ is the elapsed time from the moment of the closest location of the two opposite polarity barycenters to flare onset in hrs.  }
\label{1table}
\end{table*}

\begin{table*}[h!]
\center
\resizebox{\textwidth}{!}{
\begin{tabular}{|c|c|cccccc|}
\cline{3-8}
\multicolumn{1}{c}{}  & \multicolumn{1}{c|}{} & \multicolumn{6}{c|}{PF}  \\
\hline                   
NOAA   AR  & Flare Intensity &  $T_{Imp}^{C}$ [h]           &  $T_{Imp}^{M}$ [h]           & Opt. Height  [km] & $S_{flare}$   &$T_{est}$ [h]& $T_{D+F}$ [h]\\
\hline 
\hline    
11430	&	X1.3	&	3.4	&	5.8	& 	1845 &$<$M5&13.4&8.6\\
\hline
11515	&	X1.1	&	8.7	&	4.3	&	585 &X&89.8&35.3	\\
\hline
11520	&	X1.4	&	16.5	&	1.1	&	360	&X&65.8&67.7\\
\hline 
\multirow{2}{*}{11890}	&	X1.1	&	2.4	&	1.2	&	180 &X&52.5&17.5	\\
	&	X1.1	&	2.5	&	1.9	&	585	&$>$M5&27.6&21.4\\
\hline
11944	&	X1.2	&	4.7	&	2.2	&	225	&X&25.7&43.4\\
\hline
12017	&	X1.0	&	7.5	&	1.9	&	1080&$>$M5&24.9&32.3	\\
\hline
12158	&	X1.6	&	2.6	&	5.5	&	225&$<$M5&42.8&33.9\\
\hline
12192&X2.0&	3.3	&	8.1	&	1530&X&	41.6&33.4\\
\hline
12297	&	X2.1	&	12.7	&	3.8	&	1305&	X&47.3&37.5\\
\hline 
\multirow{3}{*}{12673}	&	X2.2/X9.3	&	6.2	&	1.7	&	1080	&X&21.4&20.6\\
	&	X1.3	&	0.3	&	1.3	&	270&X	&17.8&25.2\\
\hline
\end{tabular}}
\caption{Same as Table~\ref{1table} but for several active regions and by using PF extrapolations only. }
\label{2table}
\end{table*}

\end{document}